\documentclass[%
 reprint,
 amsmath,amssymb,
 aps,
]{revtex4-2}

\usepackage[mathlines]{lineno}
\usepackage[normalem]{ulem}
\usepackage{lmodern}
\usepackage{microtype}
\usepackage{graphicx}
\usepackage{dcolumn}
\usepackage{bm}
\usepackage{xcolor}
\usepackage{float}
\usepackage{hyperref}

\begin{document}

\preprint{APS/123-QED}

\title{Direct cross-section measurement of the weak $r$-process $^{88}$Sr($\alpha,n$)$^{91}$Zr reaction in $\nu$-driven winds of core collapse supernovae}
\author{C. Fougères$^{1}$}\thanks{Present address: CEA DAM DIF, F-91297 Arpajon, France}\email{chloe.fougeres@cea.fr}
 \author{M. L. Avila$^{1}$}
 \author{H. Jayatissa$^{1,2}$}
  \author{D. Santiago-Gonzalez$^1$}
 \author{K. Brandenburg$^3$}
\author{Z. Meisel$^3$}
  \author{P. Mohr$^4$}
  \author{F. Montes$^5$}
   \author{C. M\"uller-Gatermann$^1$}
   \author{D. Neto$^6$}
   \author{W.-J. Ong$^7$}
   \author{J. Pereira$^5$}
 \author{K. E. Rehm$^1$}
    \author{T. L. Tang$^{1,8}$}
       \author{ I. A. Tolstukhin$^1$}
         \author{L. Varriano$^9$}
  \author{G. Wilson$^{1,10}$}
  \author{J. Wu$^{1,11}$}
\affiliation{$^1$Physics Division, Argonne National Laboratory, Lemont, IL 60439, USA}
 \affiliation{$^2$Physics Division, Los Alamos National Laboratory, NM 87545, USA}
  \affiliation{$^3$Institute of Nuclear and Particle Physics, Ohio University, Athens, OH 45701, USA}%
   \affiliation{$^4$HUN-REN Institute for Nuclear Research (ATOMKI), H-4001 Debrecen, Hungary}%
  \affiliation{$^5$Facility for Rare Isotope Beams (FRIB), Michigan State University, East Lansing, MI 48824, USA}%
 \affiliation{$^6$Department of Physics, University of Illinois Chicago, 845 W. Taylor St., Chicago, IL 60607, USA}%
 \affiliation{$^7$Nuclear and Chemical Sciences Division, Lawrence Livermore National Laboratory, Livermore, CA 94550, USA}
 \affiliation{$^8$Physics Department, Florida State University, Tallahassee, FL 32306, USA}
 \affiliation{$^9$ Department of Physics, University of Chicago, Chicago, IL 60637, USA}%
\affiliation{$^{10}$Department of Physics and Astronomy, Louisiana State University, Baton Rouge, LA 70803, USA}%
 \affiliation{$^{11}$National Nuclear Data Center, Brookhaven National Laboratory, Upton, NY 11973, USA}

\date{\today}
\begin{abstract}
About half of the heavy elements beyond iron are known to be produced by the rapid neutron capture process, known as $r$-process. However, the astrophysical site producing the $r$-process is still uncertain. Chemical abundances observed in several cosmic sites indicate that different mechanisms should be at play. For instance, the abundances around silver measured in a subset of metal-poor stars indicate the presence of a weak $r$-process. This process may be active in neutrino-driven winds of core collapse supernovae where $(\alpha,n)$ reactions dominate the synthesis of $Z\sim40$ elements in the expelled materials. Scarcely measured, the rates of $(\alpha,n)$ reactions are determined from statistical Hauser-Feshbach calculations with $\alpha$-optical-model potentials, which are still poorly constrained. The uncertainties of the $(\alpha,n)$ reaction rates therefore make a significant contribution to the uncertainties of the abundances determined from stellar modeling. In this work, the $^{88}$Sr($\alpha,n)^{91}$Zr reaction which impacts the weak $r$-process abundances has been probed at astrophysics energy for the first time;~directly measuring the total cross sections at astrophysical energies of 8.37 -- 13.09~MeV in the center of mass (3.8~-- 7.5~GK). Two measurements were performed at ATLAS with the electrically-segmented ionization chamber MUSIC, in inverse kinematics, while following the active target technique. The cross sections of this $\alpha$-induced reaction on $^{88}$Sr, located at the shell closure $N=50$, have been found to be lower than expected, by a factor of 3, despite recent statistical calculations validated by measurements on neighboring nuclei. This result encourages more experimental investigations of ($\alpha,n$) reactions, at $N=50$ and towards the neutron-rich side, to further test the predictive power and reliability of such calculations.
\end{abstract}

\maketitle
\section{\label{sec:intro}Introduction}
Above helium, nuclei are produced in various stellar environments where the ongoing nucleosynthesis sheds light on the stellar conditions. Around iron, nuclear fusion ceases to contribute and a handful of other nuclear processes drive the production toward heavier masses. With respect to solar-system abundances, these elements are mainly produced in the slow and rapid neutron capture processes known as the $s$-process and $r$-process~\cite{NatureRevRprocc, RevSprocess}. Other mechanisms contribute to a smaller extend, mainly the intermediate neutron capture process (the $i$-process)~\cite{1977ApJ...212..149C}, the photodisintegration process ($p$-process)~\cite{ARNOULD20031}, and the neutrino-induced process ($\nu p$-process)~\cite{PhysRevLett.96.142502}. The exact sites where the $r$-process is active have not been fully determined yet, despite recent observational evidence in the kilonova following a binary neutron star merger (NSM) suggesting that these events are the dominant source of $r$-process elements~\cite{Kasen17,Smartt17,NatSrkilonova}. Additional sites are still being considered since some enhanced elemental abundances remain unexplained, for example lighter heavy elements ($Z=38–47$) around the first $r$-process peak~\cite{lighAbund} and actinides ($Z=90-92$) observed in {\itshape{actinide-boost}} stars~\cite{Eichler_2019, Holmbeck_2019}. Hence, more sites, such as magnetorotationally-driven supernovae~\cite{Winteler_2012,Nishimura, 2021MNRAS.501.5733R}, Collapsars~\cite{2019Natur.569..241S}, and neutrino($\nu$)-driven winds in core-collapse supernovae (CCSNe)~\cite{Hansen14,Horowitz19}, are being considered. The abundances patterns observed in a subset of metal-poor stars (see Table 2 of Ref.~\cite{Psaltis2022}) indicate the presence of a weak $r$-process ~\cite{Montes_2007, Arcones11,  QIAN2007237} producing lighter heavy elements up to silver. This is also referred as the alpha-process~\cite{1992ApJ...399..656M}.

The weak $r$-process is expected to occur in $\nu$-driven winds of CCSNe and/or NSMs~\cite{Hansen14,Horowitz19}. In these extreme winds, as matter expands and cools down, the $(\alpha,n)$ reactions are predicted to drive the synthesis of elements around $Z\sim40$ at temperatures of 2 -- 5~GK~\cite{Bliss} since they are the fastest reactions to fall out of equilibrium. At lower temperatures ($p,n$) reaction rate becomes faster. Recent sensitivity studies~\cite{Bliss2018, bliss2020nuclear, Psaltis2022} have shown that model uncertainties are presently too vital to gain insights on the stellar wind conditions, like the expansion time scale or the electron fraction, while comparing model predictions to observed Sr-to-Pd abundances in metal-poor stars. The dominant source of uncertainties from the nuclear physics side are the rates of ($\alpha,n$) reactions. These reactions are poorly measured, and rates are calculated with the statistical Hauser-Feshbach (HF) model which can lead to uncertainties of several orders of magnitude~\cite{Bliss2018, bliss2020nuclear, Psaltis2022}.

The use of the HF statistical model for $\alpha$-induced reactions is justified on these intermediate-to-heavy mass nuclei near the neutron closed shell $N=50$ given the high stellar temperatures involved ($T>1~GK$), as shown in Fig.~8 of Ref.~\cite{PhysRevC.56.1613}. The HF statistical model is based on the assumption that the reaction is a two independent step process, i.e.~the formation of the compound nucleus and its later decay by emission of $\gamma$ rays and particles. The former is characterized by the transmission coefficient of the $\alpha$ in the target nucleus ($T_{\alpha,0}$), and the latter for the ($\alpha,n$) case by the ratio of the transmission coefficient for the neutron channel ($T_n$) over the sum of the coefficients for all opened channels (${\sum T_i}$) in the compound nucleus. Hence, the cross section can be simply expressed as $\sigma(\alpha,n) \sim  T_{\alpha,0} \frac{T_n}{\sum T_i}$. At astrophysical and higher energies, well above the neutron emission threshold, $T_n$ dominates and, so, $\sigma(\alpha,n)\sim T_{\alpha,0}$ depends solely on the $\alpha$-Optical-Model Potential ($\alpha$-OMP). More details can be found in Ref.~\cite{atomkiv2}. Nevertheless, the choice of the available $\alpha$-OMPs varies the obtained ($\alpha,n$) rate within a factor of 10 -- 100. Therefore, the measurement of these $\alpha$-induced reactions relevant for the weak $r$-process is also informative for nuclear reaction theory. Probing the evolution of $(\alpha,n)$ cross sections along isotopic chains with both even–even and even–odd nuclei probed, including the $N = 50$ shell closure, would allow us to constrain nuclear properties related to $\alpha$-OMPs close to stability as well as away towards the neutron-rich side.\\

Experimentally constraining work on relevant weak $r$-process reactions has already commenced, e.g.~the measurement of the $^{100}$Mo($\alpha,xn$) reaction~\cite{Ong22,PhysRevC.104.035804} and the $^{96}$Zr($\alpha,1n$) reaction~\cite{Kiss_2021}. Nonetheless, many reactions remain to be measured~\cite{Bliss2018, bliss2020nuclear, Psaltis2022}, primarily due to the necessity for neutron-rich beams. The present radioactive beam facilities are now enabling experimental programs for weak $r$-process research at astrophysical energies ($\sim$1--3~MeV/u), thanks to the high intensities available for neutron-rich nuclei involved in this process.

Among important weak $r$-process ($\alpha,xn$) reactions, the $^{88}$Sr($\alpha,xn$)$^{91}$Zr reaction has not been measured to date. Uncertainties on its rate impact the $Z=41, 42, 44$ abundances by factors of 2--3, 5--8, and 8, respectively, in a handful of CCSNe $\nu$-driven winds conditions, according to Ref.~\cite{bliss2020nuclear, Psaltis2022}. This work presents the first measurement of $^{88}$Sr($\alpha,xn$)$^{91}$Zr cross sections performed down to astrophysical energy of 8.4~MeV in the center of mass (T$\sim$3.8~GK). The measured excitation function was compared with calculations using statistical HF models and the thermonuclear reaction rate in CCSNe $\nu$-driven winds was determined.
\section{\label{sec:exp}Experiment implementation}
\subsection{Set-up}
The measurement of the inclusive $^{88}$Sr($\alpha,xn$)$^{91}$Zr reaction cross section was carried out in inverse kinematics using the active target technique with the electrically-segmented MUlti Sampling Ionization Chamber (MUSIC)~\cite{MUSICsetup}. This technique presents many advantages for directly measuring such ($\alpha,xn$) cross sections. With an increased target thickness and a detection efficiency of $\sim$100~\%, the excitation function is probed at different center-of-mass energies while the incident mono-energetic beam is slowing down in the gaseous volume. The center-of-mass energies reached and associated resolutions are governed by the beam energy, the gas pressure, and the detector segmentation; the first two also determine the smallest measurable cross section. Additionally, such detectors are self-normalized since the incident beam is also measured. Finally, energy losses of nuclei in matter, varying as $Z^2$, allow for a clear identification of occurring ($\alpha,xn$) reactions ($Z+2$ change) in the MUSIC detector. The study of ($\alpha,xn$) reactions with MUSIC has been already proven feasible and successful~\cite{Ong22, AvilaNIM16, Avila16}.

\begin{figure}[h]
\begin{center}
\includegraphics[scale=0.38]{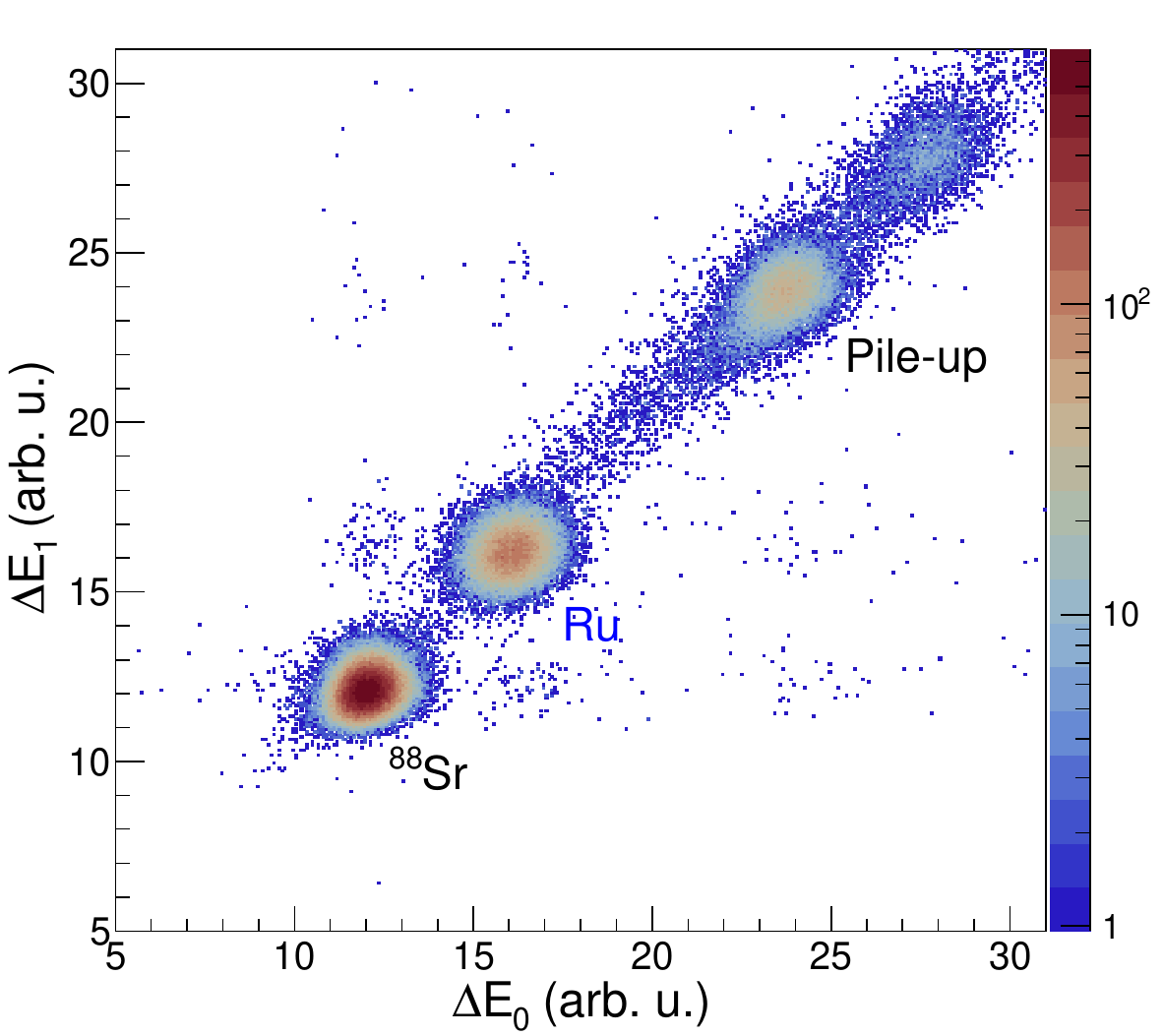}
\end{center}
\caption{Beam identification in MUSIC. Normalized energy losses in the first ($\Delta E_0$) and second  ($\Delta E_1$) strips point out the $^{88}$Sr beam spot, a Ru contaminant and pile-up events.\label{fig:beam} }
\end{figure}

The $^{88}$Sr($\alpha,xn$) reaction was assessed in two separate experiments conducted under similar experimental conditions to ensure and validate consistency in the results. The $^{88}$Sr stable beam delivered by the Argonne Tandem Linac Accelerator System (ATLAS) was selected for the charge state 16$^+$, an energy of 4.56(3)~MeV/u, and an average intensity of 3$\times10^4$~pps. The beam rate was kept at this low rate to ensure the stability of the data acquisition system. Similarly, the $^{88}$Sr beam of the second measurement was 4.55(1)~MeV/u at 3$\times10^4$~pps. The MUSIC detector, described in Ref.~\cite{MUSICsetup}, was filled with a pure $^4$He gas at 501~Torr (506~Torr for the second measurement). Two Ti foils of  1.30(5)~mg/cm$^2$ thickness located at the entrance and exit sides of the detector were used to hold the gas. The beam energy loss after passing through the entrance foil was measured to be 45.1(4)~MeV. The anode of MUSIC is segmented into 18 strips of equal width (1.578~cm) along the beam axis, the 16 inner strips are alternatively divided into a short and long section  perpendicular to the beam direction (see Fig.~1 of Ref.~\cite{MUSICsetup}). The beam composition for the first experiment is illustrated in Fig~\ref{fig:beam}
 which shows the normalized energy losses in the first two strips of the detector. The centroids of the $^{88}$Sr beam energy loss distribution for all of the strips of the detector are normalized to the energy loss of the beam in strip 0, which is estimated to be 12~MeV. A $^{99}$Ru$^{18^+}$ contaminant with an intensity of about a factor of 10 lower than the $^{88}$Sr$^{16^+}$ beam was identified based on magnetic rigidity and its energy loss profile. The $\sim$10~\% pile-up was expected for such a gaseous detector operated with rates of tens of kHz. The $^{99}$Ru contaminant was not present in the second experiment.

A new digital data acquisition system, consisting of three 1725S CAEN$^\copyright$ digitizers, was used for these two experiments which allowed to take data at higher rates (by a factor of $\sim$8) in comparison with the previous analog electronics. To ensure that the new digital system was working properly, the $^{88}$Sr($\alpha,xn$)$^{91}$Zr cross sections were also measured with the analog electronic system for a short period of time and at lower rates during the first experiment. The measured $^{88}$Sr($\alpha,xn$)$^{91}$Zr cross sections were observed to be in excellent agreement between the new digital data acquisition system and the one with analog electronics. Both data acquisition systems were triggered by the Frisch grid signal which was $\sim$10~$\mu$s faster than the signals from the anode strips. The Frisch grid is located in front of the anode pad~\cite{MUSICsetup}. 
\subsection{Energy loss measurement}
Measurement of the $^{88}$Sr beam energy losses in the windows (Ti) and gas ($^4$He) is essential considering the large discrepancies ($>$10~\%) observed at low energies when calculated using different stopping power tables from standard libraries~\cite{ATIMA,srim}. This measurement was performed with a depleted silicon surface-barrier detector which was mounted downstream of MUSIC. Due to the known pulse-height defects for heavy ions in silicon surface-barrier detectors~\cite{WILKINS1971381}, the energy response function of the detector was obtained in-beam, i.e.~by measuring several energies of $^{88}$Sr that are between the lowest (exit) and highest (incident) energy values expected during the $(\alpha,n)$ measurement. These beam energy values used for the calibration of the Si detector are:~50, 100, 200, 300 and 400 MeV. Energy corrections were implemented to take into account the dead layer at the entrance of the detector, i.e.~a thin aluminium window which has an equivalent Si thickness of 8$\times10^{-2}$~$\mu$m. Indeed, energy losses of such a heavy ion in the dead layer are not negligible:~of 0.6 -- 0.9~MeV within the range of interest. Energies were then measured after the beam travelling through ({\itshape{i}}) the Ti foils without any gas in the MUSIC chamber, ({\itshape{ii}}) the Ti foils and the $^4$He gas at 7 different pressures. The measured energy loss of $^{88}$Sr in the Ti foils was found to be in perfect agreement with the predicted value while using the stopping power table of ATIMA 1.2~\cite{ATIMA}. For the different gas pressures, the measured values were properly reproduced by considering mean values of the stopping powers from the ATIMA table using \texttt{LISE++}~\cite{lise} and the Ziegler tables using \texttt{SRIM}~\cite{srim}. An overall (mean) absolute variation of 0.62~\% was observed between measurements and Monte Carlo calculations which were performed to estimate the expected energy losses. Note that these calculations included the energy straggling of the beam in the foils and in the gas. The beam energy loss per individual strip of MUSIC was therefore measured to be 11--13~MeV. 
\subsection{Events identification}
At the probed center-of-mass energies (see Table~\ref{tab:cs}), Rutherford scattering is the dominant mechanism with cross sections higher than 1~barn according to estimations with the {\texttt{LISE++}}. Rutherford scattering and elastic/inelastic scattering cannot be distinguished from one another and all of them combined are referred as scattering or ($\alpha,\alpha'$) in this work. The ($\alpha,1n$) channel with Q$_{value}=-5.67$~MeV and ($\alpha,2n$) with Q$_{value}=-12.87$~MeV are energetically allowed as well as the ($\alpha,p$) and ($\alpha, \gamma$) channels with Q$_{value}=-6.43$~MeV and +2.96~MeV, respectively. Monte Carlo simulations of the experiment indicated that ($\alpha,p$) and ($\alpha,\gamma$) events would be poorly distinguished from the aimed ($\alpha,xn$) events. However, the calculated cross sections for ($\alpha,xn$) with the {\texttt{Talys}} code~\cite{talys1,talys2} are expected to be higher than the ($\alpha,p$) and ($\alpha, \gamma$) channels by several orders of magnitude, i.e.~a factor of $\geq$10$^4$, thereby negligible compared to the statistical uncertainty ($\geq$2~\%) of the measurement. The ($\alpha,2n$) channel is opened at the highest energy presently investigated; however, its contribution with respect to ($\alpha,1n$) is less than 5~\%, as will be shown later. Henceforth, we will refer to ($\alpha,xn$) as simply ($\alpha,n$).

The search for ($\alpha,n$) events which occur in an individual strip relies on measured energy loss $\Delta E$ {\itshape{(i)}} integrated in the “$\Sigma\Delta$E-$\Sigma_{total}\Delta$E" method, and/or {\itshape{(ii)}} local in the “energy trajectories" method. Both approaches allowed for the separation of ($\alpha,n$) events from the beam and other $\alpha$-induced reactions. The $^{88}$Sr beam was prior selected at the entrance (see Fig~\ref{fig:beam}), and, so, separated from the contaminant for the first experiment. A condition of a sharp variation (positive) in $\Delta E$ was imposed at the considered strip in order to get rid of beam-like events and of a fraction of scattering events. In the first method, the parameter $\Sigma\Delta E$ was calculated as the sum of normalized $\Delta E$ measured in the strips between the reaction position and strip 10, the latter being selected since it lies just before the Bragg peak of the $^{91}$Zr recoil. The parameter $\Sigma_{total}\Delta E$ represents the total energy deposited (or energy loss) over all segmented strips.

\begin{figure}[h]
\begin{center}
\includegraphics[scale=0.38]{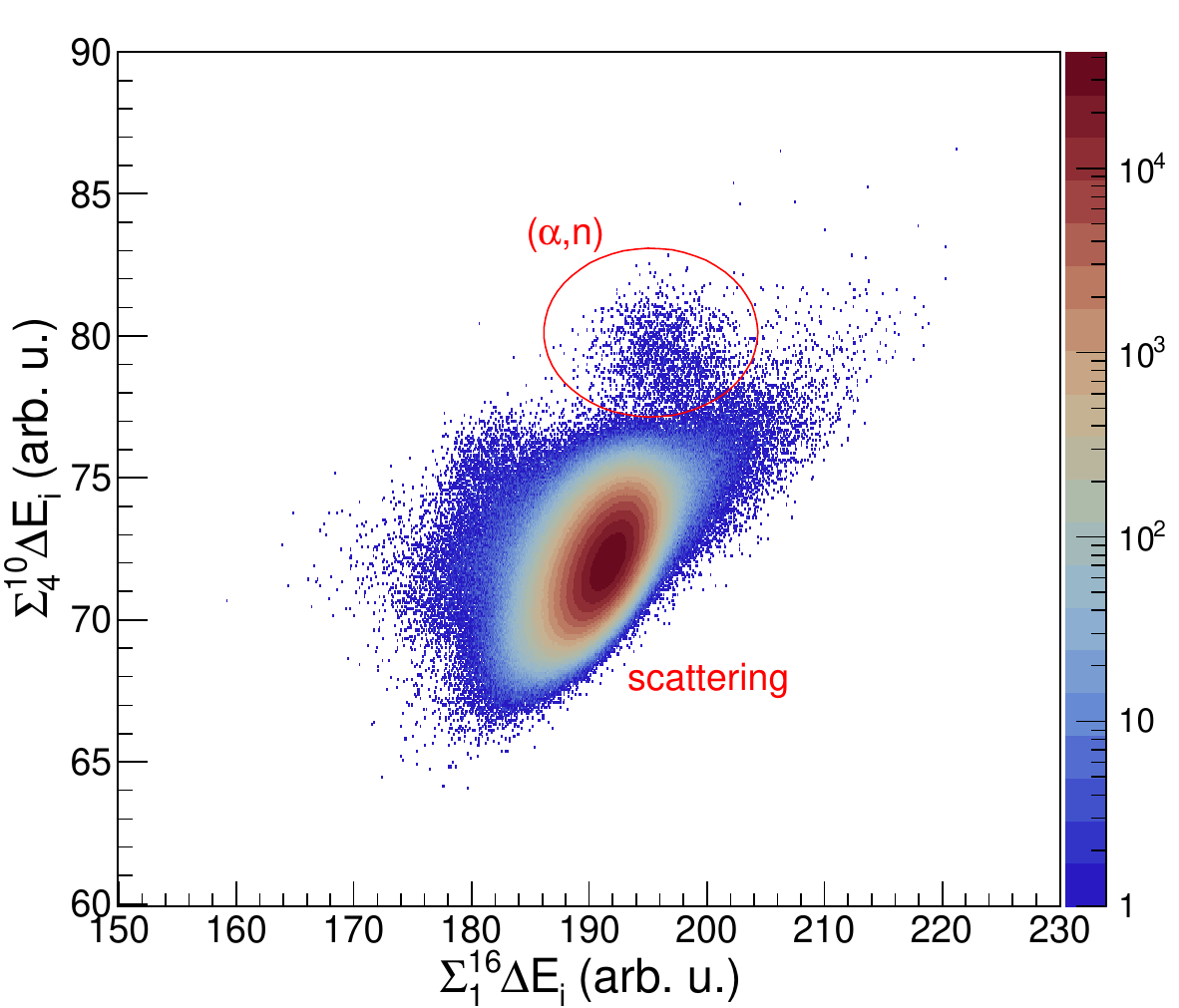}
\end{center}
\caption{Sums of normalized energy losses ($\sum_{1}^{16}\Delta E$, $\sum_{4}^{10}\Delta E$) are shown for events, selected on the entrance beam, that present a sharp increase of $\Delta E$ in strip 3. The isolated region (red circle) at higher $\sum_{4}^{10}\Delta E\sim80$ corresponds to ($\alpha,n$) events, scattering events are noticed below.}\label{fig:dEE}
\end{figure}

\begin{figure}[h]
\includegraphics[scale=0.5]{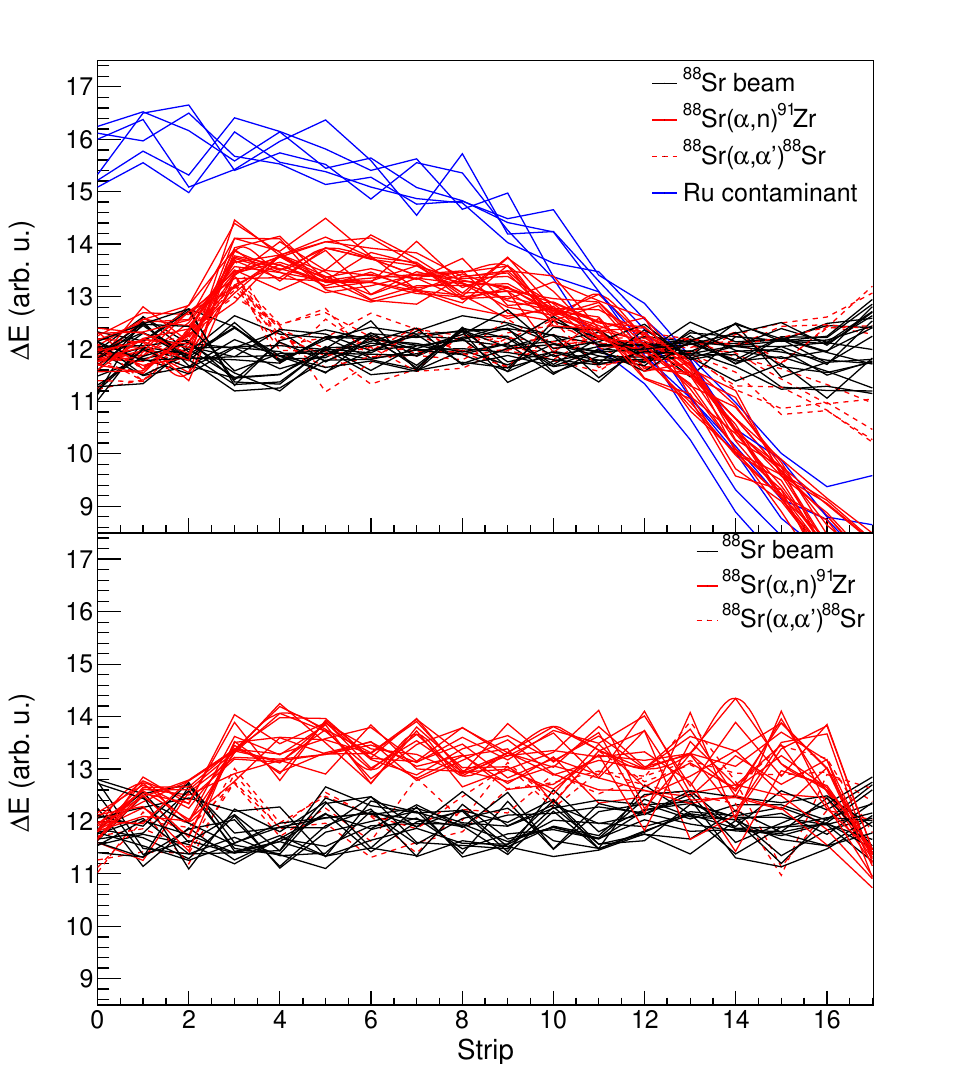}
\caption{Energy loss $\Delta E$ is shown as a function of the strip number in MUSIC for the individual trajectories associated to the unreacted beam (black curves), the ($\alpha,n$) (full red curves) and ($\alpha,\alpha'$) (dotted red curves) reactions occurring in the strip~3, for the first experiment (upper panel) and the second experiment (lower panel). The measured $^{88}$Sr($\alpha,n)^{91}$Zr events exhibit a persistent increase in $\Delta E$ from the strip~3:~the trajectories of the $^{91}$Zr recoil ($Z=40$) are observed higher than the $^{88}$Sr beam ($Z=38$). On the contrary, scattering reactions display a similar increase in $\Delta E$ but their trajectories then rejoin the beam. In the first experiment, trajectories associated with the Ru contaminant ($Z=44$) are also shown (blue curve). The $^{91}$Zr trajectories were observed lower than this higher $Z$ contaminant, strengthening the events identification. Differences in the beam Bragg's peak position and in the energy loss of the $^{91}$Zr recoil are observed between the two measurements, see details in text.\label{fig:traces}}
\end{figure}

Results in Fig~\ref{fig:dEE}
 are shown for the event identification in strip 3 of the first experiment:~the two energy regions, ($\alpha,n)$ and scattering (lower ones), do not appear fully separated. Therefore, the additional “energy trajectories" method  was implemented. This has previously been employed for  unambiguously identifying ($\alpha,n)$ events among scattering reactions~\cite{Ong22, AvilaNIM16, Avila16}. The Fig~\ref{fig:traces}  
 shows a set of trajectories, with normalized $\Delta E$, that includes the unreacted beam (black curves), the Ru contaminant (blue curves), the scattering reactions at low angles (dotted red curves) and ($\alpha,n$) of interest (full red curves) for the first experiment (upper panel) and the second one (lower panel). Note that scattering reactions at high angles resulted in a higher variation of $\Delta E$ (see also Ref.~\cite{Ong22}).

Comparison of the energy losses of the $^{91}$Zr recoil show variations between the two measurements (see Fig~\ref{fig:traces}). Indeed, their trajectories intersected with the beam trajectories around strip 12--14 for the first experiment while the intersection in the second measurement is around strip 15--16. It is observed that calculations of the expected energy losses of $^{88}$Sr and $^{91}$Zr in a pure $^4$He gas at 500~Torr agree with the second measurement (Fig~\ref{fig:traces}, lower panel). Therefore, it is inferred that an “issue” originated during the first experiment. These observed differences may originate from either a lower beam energy at the MUSIC entrance or higher energy loss in the target medium. The beam energy was accurately measured by two independent methods by the ATLAS facility, as will be discussed later. Additionally, the same Ti foils at the entrance of MUSIC were used for the two experiments, making a lower beam energy unlikely. Regarding the higher energy loss in the target medium, this can be caused by a difference in pressure or different composition of the gas. Monte Carlo simulations showed that energy losses of $^{88}$Sr, $^{91}$Zr, and $^{99}$Ru in $^4$He gas at 600 Torr agree more with the observed trajectories of the first experiment (Fig~\ref{fig:traces}, upper panel). However, the gas pressure, equivalent to the target density, was monitored every hour throughout each experiment. Since the gas used in both experiments was $^4$He of ultra high purity, a difference in composition can only be explained by a small leakage in the tube from the gas handling system to the MUSIC detector, which would have been unnoticed. Using stopping powers of $^4$He gas contaminated by a small (3(1)~\%) amount of air was found to well reproduce the observed energy losses in the first experiment. Therefore, a gas composition of $^{4}$He and air at 97:3 was considered in the first experiment.

In the second measurement, the energy resolution was poorer than the first time, primarily due to noise in the electronics of the detector (e.g. $FWHM=16\%$ against $9\%$ in strip 2), with a significant noise level observed for the energy loss in the strips (see Fig~\ref{fig:traces}, lower panel). This results in higher systematic uncertainties in the measured cross sections and prevents the extraction of events of interest for strips higher than 5. Moreover, this experiment had a shorter duration, resulting in higher statistical uncertainties.
\section{\label{sec:res}Experiment results}
\subsection{Measured cross sections}
The angle- and energy- integrated cross sections of the $^{88}$Sr($\alpha,n$)$^{91}$Zr reaction measured at center-of-mass energies of 8.37 -- 13.09~MeV are presented in Table~\ref{tab:cs}. The effective center-of-mass energies $E_{c.m., eff}$, after the thick-target yield correction~\cite{PhysRevC.104.035804}, were deduced from the measurement of the beam energy losses in the MUSIC detector. Their reported uncertainties include the uncertainties of this measurement and of the incident beam energy, corresponding to 2 -- 0.7~\% and 4 -- 2~\%, respectively. In the first measurement, the uncertainties both on the air contamination ($3^{+1}_{-3}$~\%) and on the gas pressure (501$^{+100}_{-1}$~Torr) were also taken into account. These contributions dominate the uncertainties on the $E_{c.m., eff}$. The given energy range corresponds to the width of the detector strip. Statistical and systematic contributions to the cross-section uncertainties are listed in the last two columns. In the first measurement, the systematic contribution is overall dominated by the uncertainty on the gas composition and pressure. The total uncertainty is taken as the quadratic sum of the two~\cite{bookPDG}. At low energies, systematic uncertainties increase since the separation between the ($\alpha,n$) and ($\alpha,\alpha'$) channels is more challenging (similarly to Ref.~\cite{Ong22}). At high energies, total cross sections were obtained in the two independent measures (see Table~\ref{tab:cs}). The results are in good agreement, i.e. within 6--9~\%.
\begin{table}[h]
\caption{Cross sections ($\sigma_{(\alpha,n)}$) of the $^{88}$Sr($\alpha,n$)$^{91}$Zr reaction, measured here, are reported along the effective center-of-mass energies  ($E_{c.m., eff}$). The latter were determined from the measurement of $^{88}$Sr energy losses in MUSIC. The energy range, covered in a individual strip, is mentioned. Statistical and systematic uncertainties of $\sigma_{(\alpha,n)}$ are given. At high energy, results are given for the two independent measurements.}\label{tab:cs}
\begin{ruledtabular}
\renewcommand{\arraystretch}{1.6}
\begin{tabular}{lc|ccc}
$E_{c.m., eff}$\footnote{At effective strip thickness corrected from the thick-target yield.} &
\textrm{Range}\footnote{From entrance to strip exit.} &\textrm{$\sigma_{(\alpha,n)}$ (mb)}
&\multicolumn{2}{c}{\textrm{Uncertainties (\%)}}\\
\multicolumn{2}{c|}{(MeV)}&  & statistical & systematic \\
\hline
12.89$^{+0.45}_{-0.69}$ & [13.11, 12.57] & 158(28) & 2.2 & 18.2 \\
13.09(11) & [13.29, 12.78]& 147$^{+107}_{-45}$ \footnote{Second measurement, see details in text.} & 7.1& 72\\
\hline
12.32$^{+0.46}_{-0.75}$ &[12.57, 12.01]& 123(20) & 2.4& 16.4  \\
12.55(11) & [12.78, 12.27]& 116$^{+66}_{-30}$$^{\texttt{c}}$ & 8.8& 56\\
\hline
11.75$^{+0.58}_{-0.78}$ &[12.01, 11.45]& 69(10) & 3.2& 14.9\\
12.00(10) & [12.27, 11.76]& 76$^{+62}_{-24}$$^{\texttt{c}}$ & 12& 82\\
\hline
11.20$^{+0.53}_{-0.90}$ &[11.45, 10.89]& 42.8(58) & 4.1 & 14.7 \\
11.45(10) &[11.76, 11.24]& 40$^{+31}_{-12}$$^{\texttt{c}}$ & 17& 76\\
\hline
10.65$^{+0.60}_{-1.1}$ &[10.89, 10.32]& 31.7(50) & 4.8&  15.3 \\
10.10$^{+0.65}_{-1.2}$ &[10.32, 9.75]& 10.2(18) & 8.9 & 16.1 \\
9.50$^{+0.70}_{-1.3}$ &[9.75, 9.18]& 3.33(75) & 15.8 & 16.2 \\
8.95$^{+0.75}_{-1.4}$ &[9.18, 8.60]& 1.67(60) & 22.4 & 28.3 \\
8.37 $^{+0.81}_{-1.5}$ &[8.60, 8.02]& 0.80(52) &  31.6 &  55.6\\
\end{tabular}
\end{ruledtabular}
\end{table}
\subsection{Discussion}
\begin{figure}[h]
\includegraphics[scale=0.48]{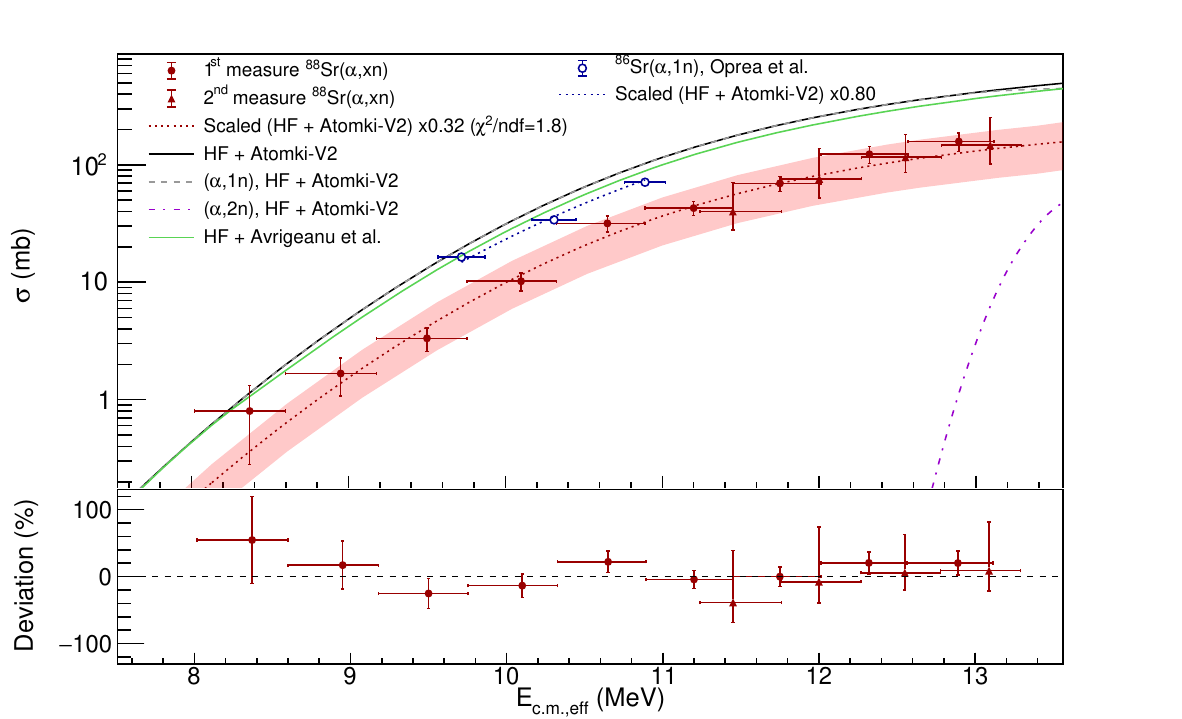}
\caption{Comparison of the measured $^{88}$Sr($\alpha,n$) cross sections to the HF statistical model along effective center-of-mass energies ($E_{c.m.,eff}$). Upper panel:~the ($\alpha,xn$) cross sections calculated with the $\alpha$-OMP of {\texttt{Atomki-V2}}~\cite{atomkiv2} (black curve) are dominated by the ($\alpha,1n$) channel (dotted grey curve), the ($\alpha,1n$) curve overlapping with the {\texttt{HF+Atomki-V2}} line. The ($\alpha,2n$) channel (dotted purple curve) starts to contribute at $E_{c.m., eff}>13$~MeV. Calculations scaled to experimental data (dotted red curve) are observed significantly lower (x0.32(9)) than theoretically predicted. They allow to access the cross sections at lower astrophysical energies (T$\leq$3.8~GK). Calculations with the $\alpha$-OMP of {\itshape{Avrigeanu et al.}}~\cite{Avrigeanu2014, Avrigeanu2023} (green curve) are also shown, resulting in a similar deviation from measures. Horizontal error bars of measured cross sections come from the width of MUSIC strips, and vertical ones are detailed in Table~\ref{tab:cs}. The colored band corresponds to 3$\sigma$ uncertainty. Measured 
$^{86}$Sr($\alpha,n$) from {\itshape{Oprea et al.}}~\cite{ref86Sr} are also given (blue points). Similar HF calculations with the {\texttt{Atomki-V2}} $\alpha$-OMP agree within a factor of 0.8 (dotted blue curve). Lower panel:~deviations of scaled calculations relative to the measure are scattered around $0\pm50$~\% and are not energy dependent.\label{fig:cs}}\end{figure}

Measurements of the $^{88}$Sr($\alpha,n$) cross sections are compared to the predictions of the HF statistical model in Fig.~\ref{fig:cs} (upper panel). Calculations were performed with the {\texttt{Talys}} code~\cite{talys1,talys2} and the $\alpha$-OMP of either {\texttt{Atomki-V2}}~\cite{atomkiv2}(black curve) or {\itshape{Avrigeanu et al.}}~\cite{Avrigeanu2014, Avrigeanu2023} (green curve). Calculations of both ($\alpha,1n$) and ($\alpha,2n$) channels (with {\texttt{Atomki-V2}} $\alpha$-OMP) are also shown. The latter is indeed open at the highest measured energy, but its contribution to the total ($\alpha,xn$) is expected to be $\sim0$ at the investigated energies. Experimental values, shown with the red points, are significantly lower than the calculated ones with both $\alpha$-OMPs. Other $\alpha$-OMPs available in {\texttt{Talys}} were found to be even more at odds with the measurements.

The {\texttt{Atomki-V2}} $\alpha$-OMP was shown to be robust and reliable with regard to measured ($\alpha,xn$) cross sections on neighboring nuclei:~~$^{86}$Sr~\cite{ref86Sr}, $^{96}$Zr~\cite{Kiss_2021}, $^{92,94}$Mo~\cite{PhysRevC.78.025804} and $^{100}$Mo~\cite{Ong22, PhysRevC.104.035804}. There, the deviation between experimental and calculated values is of a factor of 0.80, 0.66, (0.51, 0.67), and 1.2, respectively. Shown with the dotted red curve in Fig.~\ref{fig:cs}, scaling the calculations to measurement leads to a factor of 0.32(9). Furthermore, scaled $^{88}$Sr($\alpha,n$) theoretical cross sections deviate from experimental data within $\pm 50$~\% and no systematic trend is noticed along $E_{c.m., eff}$, see Fig.~\ref{fig:cs} (lower panel). The observed discrepancy between the statistical model and present measures is therefore independent of energy. This indicates that the energy dependence of the experimental data is well reproduced by the calculation.

Cross-section measurements of $\alpha$-induced reactions are direct, efficient, and accurate with the active target MUSIC, see~\cite{AvilaNIM16,Avila16,Ong22,PhysRevC.105.L042802}. The beam energy losses in gas and windows were measured. The incident beam energy was measured by two independent ways:~$(i)$ the averaged time-of-flight between three resonator pairs located upstream of MUSIC, and $(ii)$ the settings (and associated magnetic rigidity) of the Bruker magnet located after the ATLAS ion source. The resulting values agree within 1.5~\%. Finally, cross sections at high energy were measured independently on three occasions, i.e.~two measures at high beam intensity (previously described) and a measure at lower beam intensity (3~kHz) and with another data acquisition system during the first experiment. The results have been found consistent.
\\

The cross section of the $\alpha$-capture on $^{88}$Sr is fully dominated by the ($\alpha,n$) channel at energies of interest ($E_{c.m.} = 8-13$~MeV), as detailed before. This implies that $^{88}$Sr($\alpha,n$) cross sections are sensitive only to the $\alpha$-OMP. The {\texttt{Atomki-V2}} $\alpha$-OMP is based on a folding procedure which requires the density distribution of the entrance nucleus $^{88}$Sr \cite{atomkiv2}. 
The {\texttt{Atomki-V2}} $\alpha$-OMP uses the theoretical densities which are provided as a part of the TALYS code \cite{talys_general}. However, it is found that the experimental charge density of $^{88}$Sr is not very well determined because three independent data sets in the compilation of de Vries {\it et al.}~\cite{DEVRIES1987495} show significant variations of the root-mean-square (rms) radius of about 0.1 fm although the individual results claim much smaller uncertainties between 0.005 fm and 0.02 fm. Typical variations from different experiments are often of the order of 0.03 fm or even below \cite{DEVRIES1987495}. The uncertainty of 0.1 fm for the density translates to a similar uncertainty of the rms radius of the folding potential which varies between 4.86 fm and 4.97 fm. The rms radius of the potential from the theoretical TALYS density almost matches the highest value from the experimental densities of 4.97 fm. In general, a wider potential with larger rms radius leads to a more attractive nuclear potential in the barrier region and thus reduces the effective barrier height, leading to larger cross sections. We have investigated the variation of the calculated cross section which results from the choice of the theoretical or the different experimental densities. For example, at 11.5 MeV the experimental cross section is about 40 mb. The calculation with the theoretical density leads to a cross section of 182 mb. The different experimental densities yield cross sections between 151 mb and 184 mb. Consequently, the uncertainty in the density of $^{88}$Sr cannot explain the deviation between the calculated and the experimental cross sections. A dramatic reduction of the rms radius of the potential by about 0.6 fm (or about $10-15$\%) would be required to fit the experimental data which is clearly far beyond the uncertainties of the density of $^{88}$Sr.

\begin{figure}[h]
\includegraphics[scale=0.47]{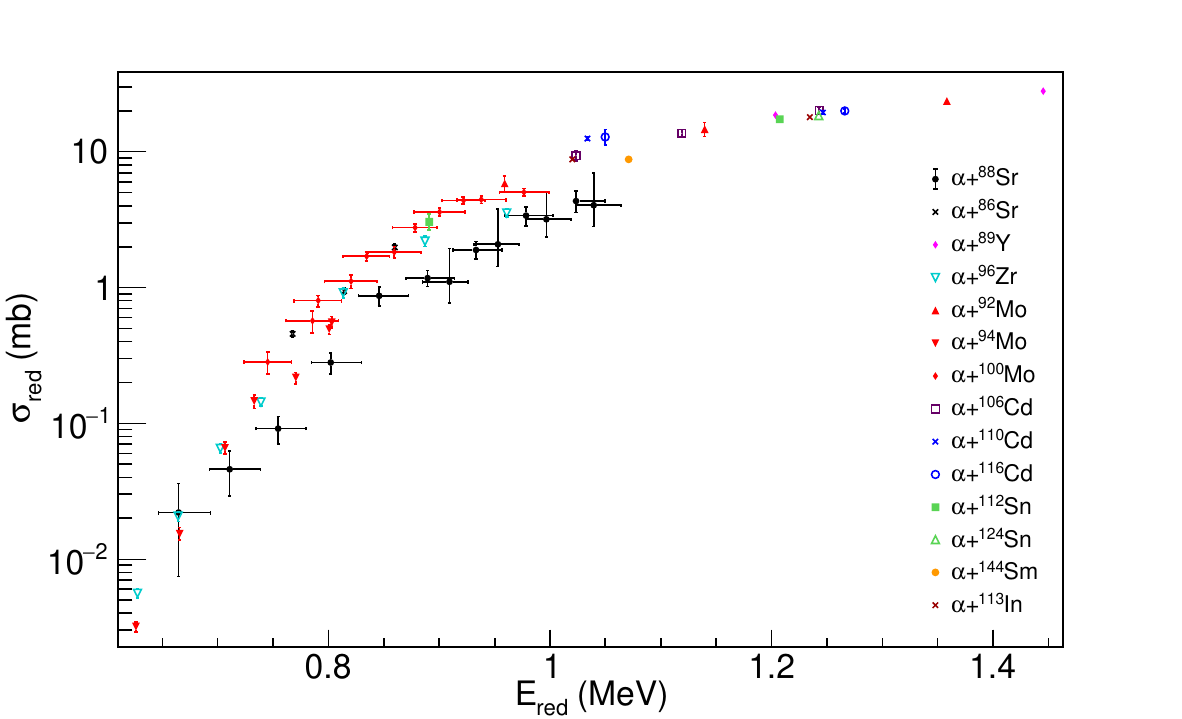}
\caption{Evolution of the reduced cross section ($\sigma_{red}$) as a function of the reduced energy ($E_{red}$) for $\alpha$-induced reactions on $A\sim90-140$ nuclei at low energy from Ref.~\cite{redSigma, ref86Sr, Kiss_2021, PhysRevC.78.025804, Ong22}. The $^{88}$Sr values from present work are outside the apparent trend which does not depend on the nucleus.\label{fig:csRed} }\end{figure}

Furthermore, the cross sections of alpha-induced reactions on $^{88}$Sr is compared to nearby systems~\cite{redSigma, ref86Sr, Kiss_2021, PhysRevC.78.025804, Ong22} with the evolution of the reduced cross section ($\sigma_{red}$) over the reduced energy ($E_{red}$) in Fig.~\ref{fig:csRed}.
 These parameters, described in~\cite{Gomes_PRC2005_reduced, redSigma}, allow to compare total cross sections of charged-particle reactions independently of the projectile, target and energy. Indeed, the reduced energy and cross section include the Coulomb barrier height and the geometrical size of the reactants system, respectively. There is a common trend, discussed in~\cite{redSigma}, from which $\alpha+^{88}$Sr unexpectedly deviates. Other nuclei located at $N=50$, $^{89}$Y and $^{92}$Mo, follow the trend. Note that a measurement of the elastic $^{88}$Sr($\alpha,\alpha'$) reaction at high (50 MeV) energy~\cite{PhysRevC.39.1281} results in $\sigma_{red}\sim53$~mb at $E_{red}=3.8$~MeV:~this fits in the observed trend but at a different (high) energy than current ones.

To date, the $^{88}$Sr($\alpha,n$)$^{91}$Zr reaction is an isolated case, exhibiting an unexpected behavior in terms of $\alpha$-induced cross sections in the $A\sim100$ mass region. This does call for further experimental investigations, particularly on Sr isotopes and towards the neutron-rich side.

\subsection{Astrophysical thermonuclear rate}
The $^{88}$Sr($\alpha,n$)$^{91}$Zr cross sections were measured at the energies associated to the Gamow temperatures of 3.8--7.5~GK. Calculations of the thermonuclear reaction rate were performed with the {\tt{EXP2RATE}} code by T.~Rauscher~\cite{refExp2Rate}. They included the cross sections from present measurement at high energies and extrapolated values at low energies (see Fig.~\ref{fig:cs}). Note that uncertainties on cross sections were deduced, if not measured, from the {\tt{HF+Atomki-V2}} calculations without scaling (upper contribution) and at 3$\sigma$ uncertainty on the scaling factor (lower contribution). The lower and upper limits of the thermonuclear reaction rate of $^{88}$Sr($\alpha,n$)$^{91}$Zr were thus obtained. They are reported in Table~\ref{tab:rate} as a function of temperature.
\begin{table}[h]
\caption{ Low, recommended and high thermonuclear rates of the $^{88}$Sr($\alpha,n$)$^{91}$Zr reaction, in units of cm$^3$ mol$^{-1}$ s$^{-1}$, as a function of temperature. \label{tab:rate} }
\begin{ruledtabular}
\renewcommand{\arraystretch}{1.5}
\begin{tabular}{lccc}
$T$ (GK)& Low & Recommended & High \\
\hline
2.0 & 5.45$\times$10$^{-12}$ & 1.04$\times$10$^{-11}$ &2.54$\times$10$^{-11}$\\
2.5 & 5.36$\times$10$^{-9}$& 9.95$\times$10$^{-9}$& 2.51$\times$10$^{-8}$\\
3.0 & 7.05$\times$10$^{-7}$ &1.30$\times$10$^{-6}$  &3.38$\times$10$^{-6}$\\
3.5 &2.85$\times$10$^{-5}$  & 5.41$\times$10$^{-5}$ &1.42$\times$10$^{-4}$\\
4.0 &5.34$\times$10$^{-4}$ & 1.05$\times$10$^{-3}$  & 2.80$\times$10$^{-3}$\\
4.5 &5.87$\times$10$^{-3}$ & 1.21$\times$10$^{-2}$& 3.19$\times$10$^{-2}$\\
5.0 & 4.37$\times$10$^{-2}$ & 9.36$\times$10$^{-2}$  & 2.42$\times$10$^{-1}$\\
5.5 &2.42$\times$10$^{-1}$ &5.34$\times$10$^{-1}$   & 1.34$\times$10$^{0}$\\
6.0 &1.06$\times$10$^{0}$ & 2.42$\times$10$^{0}$  &5.85$\times$10$^{0}$\\
6.5 &3.88$\times$10$^{0}$ &8.97$\times$10$^{0}$  &2.09$\times$10$^{1}$\\
7.0 &1.20$\times$10$^{1}$ &2.82$\times$10$^{1}$  &6.36$\times$10$^{1}$\\
7.5 &3.29$\times$10$^{1}$ &7.73$\times$10$^{1}$  &1.69$\times$10$^{2}$\\
\end{tabular}
\end{ruledtabular}
\end{table}
The rate was evaluated with the geometric mean. It was also determined from {\tt{HF+Atomki-V2}} calculations performed with the {\tt{TALYS}} code. The resulting rate was scaled by a factor of 0.32 as obtained for the cross sections. Within temperatures of 2--8~GK, these two evaluated rates agree within a factor of 0.9 -- 1.5. Combining the two, the present recommended rate is taken as the average values being also given in Table~\ref{tab:rate}.
It is worth mentioning that the $^{88}$Sr($\alpha,n$)$^{91}$Zr reaction rate becomes lower than the $^{88}$Sr($\alpha,\gamma$) and  $^{88}$Sr($\alpha,2n$) reaction rates at $T<2.3$~GK and $T>8.5$~GK, respectively.

\begin{figure}[h]
\includegraphics[scale=0.47]{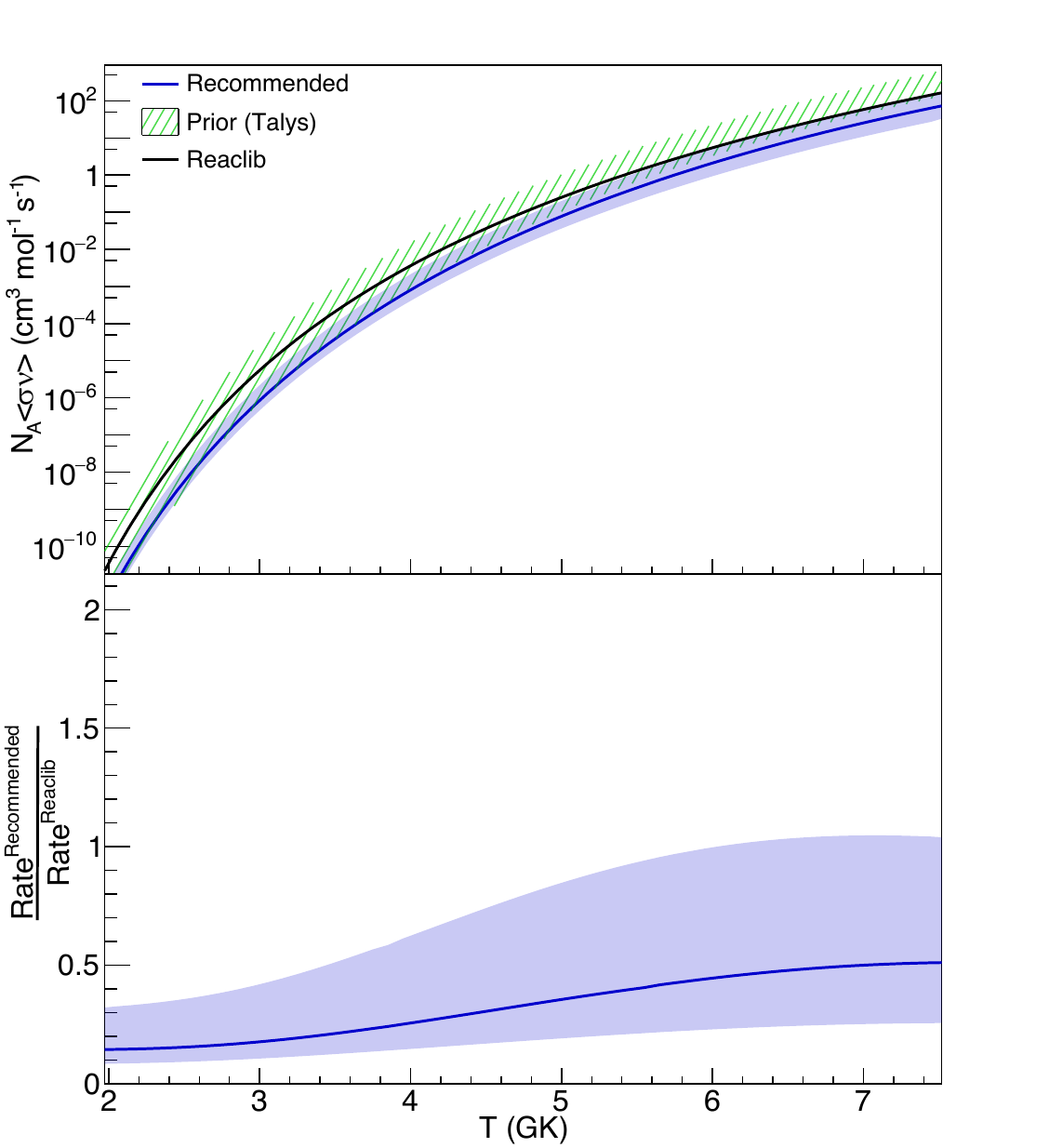}
\caption{Evolution of the thermonuclear reaction rate of $^{88}$Sr($\alpha,n$)$^{91}$Zr as a function of temperature. Upper panel:~the recommended reaction rate (blue curve) was determined from calculations with (\texttt{EXP2RATE}, \texttt{Talys+Atomki-V2}). Colored bands correspond to low and high rates (Table~\ref{tab:rate}). The prior rate band, shown as the hatched green band, was obtained with \texttt{Talys} calculations including all $\alpha$-OMPs. The rate from \texttt{ReaclibV2.2}~\cite{Cyburt2010} (black curve) is also presented. Lower panel:~the ratio of the recommended reaction rate to the rate from \texttt{ReaclibV2.2} is shown.\label{fig:rate}}\end{figure}

The evolution of the recommended reaction rate of $^{88}$Sr($\alpha,n$)$^{91}$Zr is also shown as a function of temperature in Fig.~\ref{fig:rate} (blue curve, upper panel). The colored band corresponds to its upper and lower limits.  At weak $r$-process temperatures (2--4~GK), they are of a factor of 2.4 -- 2.6. The status prior to the measurement, shown with the hatched green band (upper panel), was evaluated from {\tt{TALYS}} calculations which included all available $\alpha$-OMPs. The increased precision for the reaction rate of $^{88}$Sr($\alpha,n$)$^{91}$Zr is apparent  when previous theoretical data are compared with new recommended values. Overall, uncertainties on the rate of the reaction $^{88}$Sr($\alpha,n$)$^{91}$Zr, based on experimental data, are now less than a factor of 3, a major improvement on the assumed factor of 10--100 for $\alpha$-induced reactions when not measured. 

The recommended reaction rate of $^{88}$Sr($\alpha,n$)$^{91}$Zr is finally compared to the rate of {\texttt{ReaclibV2.2}}~\cite{Cyburt2010} in Fig.~\ref{fig:rate} (lower panel). The former is up to five times smaller than the latter at the temperatures of interest for the weak $r$-process in the CCSNe $\nu$-driven winds. 
\section{Conclusion}
The $^{88}$Sr($\alpha,n$)$^{91}$Zr reaction was reported to impact the $Z=41, 42, 44$ abundances in the weak $r$-process which are produced  in the $\nu$-driven winds after core collapse supernovae~\cite{bliss2020nuclear, Psaltis2022}. The reaction has been experimentally investigated for the first time at astrophysical temperatures of 3.8--7.5~GK by means of the active target MUSIC. Total cross sections of this reaction were directly measured from 8.37 to 13.09~MeV in the center of mass. The thermonuclear reaction rate of $^{88}$Sr($\alpha,n$)$^{91}$Zr has been determined at weak $r$-process temperatures with significantly improved uncertainties:~up to a factor of 3 presently against a factor of 100 prior to this study. 

When they have not been measured, the ($\alpha,n$) cases impacting the weak $r$-process carry important uncertainties due to the choice of the $\alpha$-OMP in the Hauser-Feshbach statistical model. The potential {\texttt{Atomki-V2}}~\cite{atomkiv2} has been proven to bring calculations in excellent agreement with measured data. Indeed, deviations between measured and calculated cross sections were observed within $\sim50$~\% for several nuclei nearby $^{88}$Sr. Present $^{88}$Sr($\alpha,n$) measurement is surprisingly found lower than calculations by a factor of 3. This singular experimental result has been shown reliable thanks to two independent measures. The $^{88}$Sr nucleus is located at the $N=50$ shell closure, near the neutron-rich side, where the nuclear level density and the nuclear deformation of the ground state may give some hints of why HF statistical model predictions disagree with experimental values. 

Nucleosynthesis calculations using reaction rates based on the HF statistical model need to be more accurate for the modeling of CCSNe $\nu$-driven winds and insightful model-to-observations comparisons of abundances in metal-poor stars~\cite{Psaltis2022}. Thereby, further tests of the predictive power of $\alpha$-OMPs around the $N=50$ shell closure and towards the neutron-rich radioisotopes are of keen interest. The experimental method based on the active target MUSIC is apropos, particularly at ATLAS and FRIB facilities where such investigations are now reachable thanks to the improved luminosities of neutron-rich beams.\\

\begin{acknowledgments}
The authors thank the support of the ATLAS beam and detection physicists. This material is based upon work supported by the U.S. Department of Energy, Office of Science, Office of Nuclear Physics, under Contract No. DE-AC02-06CH11357. This research used resources of Argonne National Laboratory’s ATLAS facility, which is a DOE Office of Science User Facility.
P.M. would like to acknowledge the support of NKFIH (grant K134197).

\end{acknowledgments}

\bibliographystyle{apsrev4-1}
\bibliography{apssamp}

\begin{thebibliography}{52}%
\makeatletter
\providecommand \@ifxundefined [1]{%
 \@ifx{#1\undefined}
}%
\providecommand \@ifnum [1]{%
 \ifnum #1\expandafter \@firstoftwo
 \else \expandafter \@secondoftwo
 \fi
}%
\providecommand \@ifx [1]{%
 \ifx #1\expandafter \@firstoftwo
 \else \expandafter \@secondoftwo
 \fi
}%
\providecommand \natexlab [1]{#1}%
\providecommand \enquote  [1]{``#1''}%
\providecommand \bibnamefont  [1]{#1}%
\providecommand \bibfnamefont [1]{#1}%
\providecommand \citenamefont [1]{#1}%
\providecommand \href@noop [0]{\@secondoftwo}%
\providecommand \href [0]{\begingroup \@sanitize@url \@href}%
\providecommand \@href[1]{\@@startlink{#1}\@@href}%
\providecommand \@@href[1]{\endgroup#1\@@endlink}%
\providecommand \@sanitize@url [0]{\catcode `\\12\catcode `\$12\catcode
  `\&12\catcode `\#12\catcode `\^12\catcode `\_12\catcode `\%12\relax}%
\providecommand \@@startlink[1]{}%
\providecommand \@@endlink[0]{}%
\providecommand \url  [0]{\begingroup\@sanitize@url \@url }%
\providecommand \@url [1]{\endgroup\@href {#1}{\urlprefix }}%
\providecommand \urlprefix  [0]{URL }%
\providecommand \Eprint [0]{\href }%
\providecommand \doibase [0]{http://dx.doi.org/}%
\providecommand \selectlanguage [0]{\@gobble}%
\providecommand \bibinfo  [0]{\@secondoftwo}%
\providecommand \bibfield  [0]{\@secondoftwo}%
\providecommand \translation [1]{[#1]}%
\providecommand \BibitemOpen [0]{}%
\providecommand \bibitemStop [0]{}%
\providecommand \bibitemNoStop [0]{.\EOS\space}%
\providecommand \EOS [0]{\spacefactor3000\relax}%
\providecommand \BibitemShut  [1]{\csname bibitem#1\endcsname}%
\let\auto@bib@innerbib\@empty
\bibitem [{\citenamefont {Siegel}(2022)}]{NatureRevRprocc}%
  \BibitemOpen
  \bibfield  {author} {\bibinfo {author} {\bibfnamefont {D.~M.}\ \bibnamefont
  {Siegel}},\ }\href {\doibase https://doi.org/10.1038/s42254-022-00439-1}
  {\bibfield  {journal} {\bibinfo  {journal} {Nature Reviews Physics}\ }\textbf
  {\bibinfo {volume} {4}},\ \bibinfo {pages} {306–318} (\bibinfo {year}
  {2022})}\BibitemShut {NoStop}%
\bibitem [{\citenamefont {K\"appeler}\ \emph {et~al.}(2011)\citenamefont
  {K\"appeler}, \citenamefont {Gallino}, \citenamefont {Bisterzo},\ and\
  \citenamefont {Aoki}}]{RevSprocess}%
  \BibitemOpen
  \bibfield  {author} {\bibinfo {author} {\bibfnamefont {F.}~\bibnamefont
  {K\"appeler}}, \bibinfo {author} {\bibfnamefont {R.}~\bibnamefont {Gallino}},
  \bibinfo {author} {\bibfnamefont {S.}~\bibnamefont {Bisterzo}}, \ and\
  \bibinfo {author} {\bibfnamefont {W.}~\bibnamefont {Aoki}},\ }\href {\doibase
  https://doi.org/10.1103/RevModPhys.83.157} {\bibfield  {journal} {\bibinfo
  {journal} {Rev. Mod. Phys.}\ }\textbf {\bibinfo {volume} {83}},\ \bibinfo
  {pages} {157} (\bibinfo {year} {2011})}\BibitemShut {NoStop}%
\bibitem [{\citenamefont {{Cowan}}\ and\ \citenamefont
  {{Rose}}(1977)}]{1977ApJ...212..149C}%
  \BibitemOpen
  \bibfield  {author} {\bibinfo {author} {\bibfnamefont {J.~J.}\ \bibnamefont
  {{Cowan}}}\ and\ \bibinfo {author} {\bibfnamefont {W.~K.}\ \bibnamefont
  {{Rose}}},\ }\href {\doibase 10.1086/155030} {\bibfield  {journal} {\bibinfo
  {journal} {\apj}\ }\textbf {\bibinfo {volume} {212}},\ \bibinfo {pages} {149}
  (\bibinfo {year} {1977})}\BibitemShut {NoStop}%
\bibitem [{\citenamefont {Arnould}\ and\ \citenamefont
  {Goriely}(2003)}]{ARNOULD20031}%
  \BibitemOpen
  \bibfield  {author} {\bibinfo {author} {\bibfnamefont {M.}~\bibnamefont
  {Arnould}}\ and\ \bibinfo {author} {\bibfnamefont {S.}~\bibnamefont
  {Goriely}},\ }\href {\doibase https://doi.org/10.1016/S0370-1573(03)00242-4}
  {\bibfield  {journal} {\bibinfo  {journal} {Physics Reports}\ }\textbf
  {\bibinfo {volume} {384}},\ \bibinfo {pages} {1} (\bibinfo {year}
  {2003})}\BibitemShut {NoStop}%
\bibitem [{\citenamefont {Fr\"ohlich}\ \emph {et~al.}(2006)\citenamefont
  {Fr\"ohlich}, \citenamefont {Mart\'{\i}nez-Pinedo}, \citenamefont
  {Liebend\"orfer}, \citenamefont {Thielemann}, \citenamefont {Bravo},
  \citenamefont {Hix}, \citenamefont {Langanke},\ and\ \citenamefont
  {Zinner}}]{PhysRevLett.96.142502}%
  \BibitemOpen
  \bibfield  {author} {\bibinfo {author} {\bibfnamefont {C.}~\bibnamefont
  {Fr\"ohlich}}, \bibinfo {author} {\bibfnamefont {G.}~\bibnamefont
  {Mart\'{\i}nez-Pinedo}}, \bibinfo {author} {\bibfnamefont {M.}~\bibnamefont
  {Liebend\"orfer}}, \bibinfo {author} {\bibfnamefont {F.-K.}\ \bibnamefont
  {Thielemann}}, \bibinfo {author} {\bibfnamefont {E.}~\bibnamefont {Bravo}},
  \bibinfo {author} {\bibfnamefont {W.~R.}\ \bibnamefont {Hix}}, \bibinfo
  {author} {\bibfnamefont {K.}~\bibnamefont {Langanke}}, \ and\ \bibinfo
  {author} {\bibfnamefont {N.~T.}\ \bibnamefont {Zinner}},\ }\href {\doibase
  https://doi.org/10.1103/PhysRevLett.96.142502} {\bibfield  {journal}
  {\bibinfo  {journal} {Phys. Rev. Lett.}\ }\textbf {\bibinfo {volume} {96}},\
  \bibinfo {pages} {142502} (\bibinfo {year} {2006})}\BibitemShut {NoStop}%
\bibitem [{\citenamefont {{Kasen}}\ \emph {et~al.}(2017)\citenamefont
  {{Kasen}}, \citenamefont {{Metzger}}, \citenamefont {{Barnes}}, \citenamefont
  {{Quataert}},\ and\ \citenamefont {{Ramirez-Ruiz}}}]{Kasen17}%
  \BibitemOpen
  \bibfield  {author} {\bibinfo {author} {\bibfnamefont {D.}~\bibnamefont
  {{Kasen}}}, \bibinfo {author} {\bibfnamefont {B.}~\bibnamefont {{Metzger}}},
  \bibinfo {author} {\bibfnamefont {J.}~\bibnamefont {{Barnes}}}, \bibinfo
  {author} {\bibfnamefont {E.}~\bibnamefont {{Quataert}}}, \ and\ \bibinfo
  {author} {\bibfnamefont {E.}~\bibnamefont {{Ramirez-Ruiz}}},\ }\href
  {\doibase 10.1038/nature24453} {\bibfield  {journal} {\bibinfo  {journal}
  {Nature}\ }\textbf {\bibinfo {volume} {551}},\ \bibinfo {pages} {80}
  (\bibinfo {year} {2017})}\BibitemShut {NoStop}%
\bibitem [{\citenamefont {Smartt}\ \emph {et~al.}(2017)\citenamefont {Smartt},
  \citenamefont {Chen}, \citenamefont {Jerkstrand} \emph {et~al.}}]{Smartt17}%
  \BibitemOpen
  \bibfield  {author} {\bibinfo {author} {\bibfnamefont {S.}~\bibnamefont
  {Smartt}}, \bibinfo {author} {\bibfnamefont {T.-W.}\ \bibnamefont {Chen}},
  \bibinfo {author} {\bibfnamefont {A.}~\bibnamefont {Jerkstrand}},  \emph
  {et~al.},\ }\href {\doibase https://doi.org/10.1038/nature24303} {\bibfield
  {journal} {\bibinfo  {journal} {Nature}\ }\textbf {\bibinfo {volume} {551}},\
  \bibinfo {pages} {75} (\bibinfo {year} {2017})}\BibitemShut {NoStop}%
\bibitem [{\citenamefont {Watson}\ \emph {et~al.}(2019)\citenamefont {Watson},
  \citenamefont {Hansen}, \citenamefont {Selsing} \emph
  {et~al.}}]{NatSrkilonova}%
  \BibitemOpen
  \bibfield  {author} {\bibinfo {author} {\bibfnamefont {D.}~\bibnamefont
  {Watson}}, \bibinfo {author} {\bibfnamefont {C.}~\bibnamefont {Hansen}},
  \bibinfo {author} {\bibfnamefont {J.}~\bibnamefont {Selsing}},  \emph
  {et~al.},\ }\href {\doibase https://doi.org/10.1038/s41586-019-1676-3}
  {\bibfield  {journal} {\bibinfo  {journal} {Nature}\ }\textbf {\bibinfo
  {volume} {574}},\ \bibinfo {pages} {497} (\bibinfo {year}
  {2019})}\BibitemShut {NoStop}%
\bibitem [{\citenamefont {Sneden}\ \emph {et~al.}(2008)\citenamefont {Sneden},
  \citenamefont {Cowan},\ and\ \citenamefont {Gallino}}]{lighAbund}%
  \BibitemOpen
  \bibfield  {author} {\bibinfo {author} {\bibfnamefont {C.}~\bibnamefont
  {Sneden}}, \bibinfo {author} {\bibfnamefont {J.~J.}\ \bibnamefont {Cowan}}, \
  and\ \bibinfo {author} {\bibfnamefont {R.}~\bibnamefont {Gallino}},\ }\href
  {\doibase 10.1146/annurev.astro.46.060407.145207} {\bibfield  {journal}
  {\bibinfo  {journal} {Annual Review of Astronomy and Astrophysics}\ }\textbf
  {\bibinfo {volume} {46}},\ \bibinfo {pages} {241} (\bibinfo {year}
  {2008})}\BibitemShut {NoStop}%
\bibitem [{\citenamefont {Eichler}\ \emph {et~al.}(2019)\citenamefont
  {Eichler}, \citenamefont {Sayar}, \citenamefont {Arcones},\ and\
  \citenamefont {Rauscher}}]{Eichler_2019}%
  \BibitemOpen
  \bibfield  {author} {\bibinfo {author} {\bibfnamefont {M.}~\bibnamefont
  {Eichler}}, \bibinfo {author} {\bibfnamefont {W.}~\bibnamefont {Sayar}},
  \bibinfo {author} {\bibfnamefont {A.}~\bibnamefont {Arcones}}, \ and\
  \bibinfo {author} {\bibfnamefont {T.}~\bibnamefont {Rauscher}},\ }\href
  {\doibase 10.3847/1538-4357/ab24cf} {\bibfield  {journal} {\bibinfo
  {journal} {\apj}\ }\textbf {\bibinfo {volume} {879}},\ \bibinfo {pages} {47}
  (\bibinfo {year} {2019})}\BibitemShut {NoStop}%
\bibitem [{\citenamefont {Holmbeck}\ \emph {et~al.}(2018)\citenamefont
  {Holmbeck}, \citenamefont {Sprouse}, \citenamefont {Mumpower}, \citenamefont
  {Vassh}, \citenamefont {Surman}, \citenamefont {Beers},\ and\ \citenamefont
  {Kawano}}]{Holmbeck_2019}%
  \BibitemOpen
  \bibfield  {author} {\bibinfo {author} {\bibfnamefont {E.~M.}\ \bibnamefont
  {Holmbeck}}, \bibinfo {author} {\bibfnamefont {T.~M.}\ \bibnamefont
  {Sprouse}}, \bibinfo {author} {\bibfnamefont {M.~R.}\ \bibnamefont
  {Mumpower}}, \bibinfo {author} {\bibfnamefont {N.}~\bibnamefont {Vassh}},
  \bibinfo {author} {\bibfnamefont {R.}~\bibnamefont {Surman}}, \bibinfo
  {author} {\bibfnamefont {T.~C.}\ \bibnamefont {Beers}}, \ and\ \bibinfo
  {author} {\bibfnamefont {T.}~\bibnamefont {Kawano}},\ }\href {\doibase
  https://doi.org/10.3847/1538-4357/aaefef} {\bibfield  {journal} {\bibinfo
  {journal} {\apj}\ }\textbf {\bibinfo {volume} {870}},\ \bibinfo {pages} {23}
  (\bibinfo {year} {2018})}\BibitemShut {NoStop}%
\bibitem [{\citenamefont {Winteler}\ \emph {et~al.}(2012)\citenamefont
  {Winteler}, \citenamefont {Käppeli}, \citenamefont {Perego}, \citenamefont
  {Arcones}, \citenamefont {Vasset}, \citenamefont {Nishimura}, \citenamefont
  {Liebendörfer},\ and\ \citenamefont {Thielemann}}]{Winteler_2012}%
  \BibitemOpen
  \bibfield  {author} {\bibinfo {author} {\bibfnamefont {C.}~\bibnamefont
  {Winteler}}, \bibinfo {author} {\bibfnamefont {R.}~\bibnamefont {Käppeli}},
  \bibinfo {author} {\bibfnamefont {A.}~\bibnamefont {Perego}}, \bibinfo
  {author} {\bibfnamefont {A.}~\bibnamefont {Arcones}}, \bibinfo {author}
  {\bibfnamefont {N.}~\bibnamefont {Vasset}}, \bibinfo {author} {\bibfnamefont
  {N.}~\bibnamefont {Nishimura}}, \bibinfo {author} {\bibfnamefont
  {M.}~\bibnamefont {Liebendörfer}}, \ and\ \bibinfo {author} {\bibfnamefont
  {F.-K.}\ \bibnamefont {Thielemann}},\ }\href {\doibase
  https://doi.org/10.1088/2041-8205/750/1/L22} {\bibfield  {journal} {\bibinfo
  {journal} {Astrophys. J. Lett.}\ }\textbf {\bibinfo {volume} {750}},\
  \bibinfo {pages} {L22} (\bibinfo {year} {2012})}\BibitemShut {NoStop}%
\bibitem [{\citenamefont {Nishimura}\ \emph {et~al.}(2017)\citenamefont
  {Nishimura}, \citenamefont {Sawai}, \citenamefont {Takiwaki}, \citenamefont
  {Yamada},\ and\ \citenamefont {Thielemann}}]{Nishimura}%
  \BibitemOpen
  \bibfield  {author} {\bibinfo {author} {\bibfnamefont {N.}~\bibnamefont
  {Nishimura}}, \bibinfo {author} {\bibfnamefont {H.}~\bibnamefont {Sawai}},
  \bibinfo {author} {\bibfnamefont {T.}~\bibnamefont {Takiwaki}}, \bibinfo
  {author} {\bibfnamefont {S.}~\bibnamefont {Yamada}}, \ and\ \bibinfo {author}
  {\bibfnamefont {F.-K.}\ \bibnamefont {Thielemann}},\ }\href {\doibase
  https://doi.org/10.3847/2041-8213/aa5dee} {\bibfield  {journal} {\bibinfo
  {journal} {Astrophys. J. Lett}\ }\textbf {\bibinfo {volume} {836}},\ \bibinfo
  {pages} {L21} (\bibinfo {year} {2017})}\BibitemShut {NoStop}%
\bibitem [{\citenamefont {{Reichert}}\ \emph {et~al.}(2021)\citenamefont
  {{Reichert}}, \citenamefont {{Obergaulinger}}, \citenamefont {{Eichler}},
  \citenamefont {{Aloy}},\ and\ \citenamefont
  {{Arcones}}}]{2021MNRAS.501.5733R}%
  \BibitemOpen
  \bibfield  {author} {\bibinfo {author} {\bibfnamefont {M.}~\bibnamefont
  {{Reichert}}}, \bibinfo {author} {\bibfnamefont {M.}~\bibnamefont
  {{Obergaulinger}}}, \bibinfo {author} {\bibfnamefont {M.}~\bibnamefont
  {{Eichler}}}, \bibinfo {author} {\bibfnamefont {M.~{\'A}.}\ \bibnamefont
  {{Aloy}}}, \ and\ \bibinfo {author} {\bibfnamefont {A.}~\bibnamefont
  {{Arcones}}},\ }\href {\doibase https://doi.org/10.1093/mnras/stab029}
  {\bibfield  {journal} {\bibinfo  {journal} {Mon. Not. R. Astron. Soc.}\
  }\textbf {\bibinfo {volume} {501}},\ \bibinfo {pages} {5733} (\bibinfo {year}
  {2021})}\BibitemShut {NoStop}%
\bibitem [{\citenamefont {{Siegel}}\ \emph {et~al.}(2019)\citenamefont
  {{Siegel}}, \citenamefont {{Barnes}},\ and\ \citenamefont
  {{Metzger}}}]{2019Natur.569..241S}%
  \BibitemOpen
  \bibfield  {author} {\bibinfo {author} {\bibfnamefont {D.~M.}\ \bibnamefont
  {{Siegel}}}, \bibinfo {author} {\bibfnamefont {J.}~\bibnamefont {{Barnes}}},
  \ and\ \bibinfo {author} {\bibfnamefont {B.~D.}\ \bibnamefont {{Metzger}}},\
  }\href {\doibase 10.1038/s41586-019-1136-0} {\bibfield  {journal} {\bibinfo
  {journal} {\nat}\ }\textbf {\bibinfo {volume} {569}},\ \bibinfo {pages} {241}
  (\bibinfo {year} {2019})}\BibitemShut {NoStop}%
\bibitem [{\citenamefont {{Hansen}}\ \emph {et~al.}(2014)\citenamefont
  {{Hansen}}, \citenamefont {{Montes}},\ and\ \citenamefont
  {{Arcones}}}]{Hansen14}%
  \BibitemOpen
  \bibfield  {author} {\bibinfo {author} {\bibfnamefont {C.~J.}\ \bibnamefont
  {{Hansen}}}, \bibinfo {author} {\bibfnamefont {F.}~\bibnamefont {{Montes}}},
  \ and\ \bibinfo {author} {\bibfnamefont {A.}~\bibnamefont {{Arcones}}},\
  }\href {\doibase 10.1088/0004-637X/797/2/123} {\bibfield  {journal} {\bibinfo
   {journal} {\apj}\ }\textbf {\bibinfo {volume} {797}},\ \bibinfo {eid} {123}
  (\bibinfo {year} {2014})}\BibitemShut {NoStop}%
\bibitem [{\citenamefont {Horowitz}\ \emph {et~al.}(2019)\citenamefont
  {Horowitz}, \citenamefont {Arcones}, \citenamefont {C{\^o}t{\'e}},
  \citenamefont {Dillmann}, \citenamefont {Nazarewicz}, \citenamefont
  {Roederer}, \citenamefont {Schatz}, \citenamefont {Aprahamian}, \citenamefont
  {Atanasov}, \citenamefont {Bauswein} \emph {et~al.}}]{Horowitz19}%
  \BibitemOpen
  \bibfield  {author} {\bibinfo {author} {\bibfnamefont {C.~J.}\ \bibnamefont
  {Horowitz}}, \bibinfo {author} {\bibfnamefont {A.}~\bibnamefont {Arcones}},
  \bibinfo {author} {\bibfnamefont {B.}~\bibnamefont {C{\^o}t{\'e}}}, \bibinfo
  {author} {\bibfnamefont {I.}~\bibnamefont {Dillmann}}, \bibinfo {author}
  {\bibfnamefont {W.}~\bibnamefont {Nazarewicz}}, \bibinfo {author}
  {\bibfnamefont {I.~U.}\ \bibnamefont {Roederer}}, \bibinfo {author}
  {\bibfnamefont {H.}~\bibnamefont {Schatz}}, \bibinfo {author} {\bibfnamefont
  {A.}~\bibnamefont {Aprahamian}}, \bibinfo {author} {\bibfnamefont
  {D.}~\bibnamefont {Atanasov}}, \bibinfo {author} {\bibfnamefont
  {A.}~\bibnamefont {Bauswein}},  \emph {et~al.},\ }\href {\doibase
  https://doi.org/10.1088/1361-6471/ab0849} {\bibfield  {journal} {\bibinfo
  {journal} {J. Phys. G: Nucl. Part. Phys.}\ }\textbf {\bibinfo {volume}
  {46}},\ \bibinfo {pages} {083001} (\bibinfo {year} {2019})}\BibitemShut
  {NoStop}%
\bibitem [{\citenamefont {Psaltis}\ \emph {et~al.}(2022)\citenamefont
  {Psaltis}, \citenamefont {Arcones}, \citenamefont {Montes}, \citenamefont
  {Mohr}, \citenamefont {Hansen}, \citenamefont {Jacobi},\ and\ \citenamefont
  {Schatz}}]{Psaltis2022}%
  \BibitemOpen
  \bibfield  {author} {\bibinfo {author} {\bibfnamefont {A.}~\bibnamefont
  {Psaltis}}, \bibinfo {author} {\bibfnamefont {A.}~\bibnamefont {Arcones}},
  \bibinfo {author} {\bibfnamefont {F.}~\bibnamefont {Montes}}, \bibinfo
  {author} {\bibfnamefont {P.}~\bibnamefont {Mohr}}, \bibinfo {author}
  {\bibfnamefont {C.~J.}\ \bibnamefont {Hansen}}, \bibinfo {author}
  {\bibfnamefont {M.}~\bibnamefont {Jacobi}}, \ and\ \bibinfo {author}
  {\bibfnamefont {H.}~\bibnamefont {Schatz}},\ }\href {\doibase
  https://doi.org/10.3847/1538-4357/ac7da7} {\bibfield  {journal} {\bibinfo
  {journal} {\apj}\ }\textbf {\bibinfo {volume} {935}},\ \bibinfo {pages} {27}
  (\bibinfo {year} {2022})}\BibitemShut {NoStop}%
\bibitem [{\citenamefont {Montes}\ \emph {et~al.}(2007)\citenamefont {Montes},
  \citenamefont {Beers}, \citenamefont {Cowan}, \citenamefont {Elliot},
  \citenamefont {Farouqi}, \citenamefont {Gallino}, \citenamefont {Heil},
  \citenamefont {Kratz}, \citenamefont {Pfeiffer}, \citenamefont {Pignatari},\
  and\ \citenamefont {Schatz}}]{Montes_2007}%
  \BibitemOpen
  \bibfield  {author} {\bibinfo {author} {\bibfnamefont {F.}~\bibnamefont
  {Montes}}, \bibinfo {author} {\bibfnamefont {T.~C.}\ \bibnamefont {Beers}},
  \bibinfo {author} {\bibfnamefont {J.}~\bibnamefont {Cowan}}, \bibinfo
  {author} {\bibfnamefont {T.}~\bibnamefont {Elliot}}, \bibinfo {author}
  {\bibfnamefont {K.}~\bibnamefont {Farouqi}}, \bibinfo {author} {\bibfnamefont
  {R.}~\bibnamefont {Gallino}}, \bibinfo {author} {\bibfnamefont
  {M.}~\bibnamefont {Heil}}, \bibinfo {author} {\bibfnamefont {K.-L.}\
  \bibnamefont {Kratz}}, \bibinfo {author} {\bibfnamefont {B.}~\bibnamefont
  {Pfeiffer}}, \bibinfo {author} {\bibfnamefont {M.}~\bibnamefont {Pignatari}},
  \ and\ \bibinfo {author} {\bibfnamefont {H.}~\bibnamefont {Schatz}},\ }\href
  {\doibase 10.1086/523084} {\bibfield  {journal} {\bibinfo  {journal} {\apj}\
  }\textbf {\bibinfo {volume} {671}},\ \bibinfo {pages} {1685} (\bibinfo {year}
  {2007})}\BibitemShut {NoStop}%
\bibitem [{\citenamefont {{Arcones}}\ and\ \citenamefont
  {{Montes}}(2011)}]{Arcones11}%
  \BibitemOpen
  \bibfield  {author} {\bibinfo {author} {\bibfnamefont {A.}~\bibnamefont
  {{Arcones}}}\ and\ \bibinfo {author} {\bibfnamefont {F.}~\bibnamefont
  {{Montes}}},\ }\href {\doibase 10.1088/0004-637X/731/1/5} {\bibfield
  {journal} {\bibinfo  {journal} {\apj}\ }\textbf {\bibinfo {volume} {731}},\
  \bibinfo {eid} {5} (\bibinfo {year} {2011})}\BibitemShut {NoStop}%
\bibitem [{\citenamefont {Qian}\ and\ \citenamefont
  {Wasserburg}(2007)}]{QIAN2007237}%
  \BibitemOpen
  \bibfield  {author} {\bibinfo {author} {\bibfnamefont {Y.-Z.}\ \bibnamefont
  {Qian}}\ and\ \bibinfo {author} {\bibfnamefont {G.}~\bibnamefont
  {Wasserburg}},\ }\href {\doibase
  https://doi.org/10.1016/j.physrep.2007.02.006} {\bibfield  {journal}
  {\bibinfo  {journal} {Physics Reports}\ }\textbf {\bibinfo {volume} {442}},\
  \bibinfo {pages} {237} (\bibinfo {year} {2007})}\BibitemShut {NoStop}%
\bibitem [{\citenamefont {{Meyer}}\ \emph {et~al.}(1992)\citenamefont
  {{Meyer}}, \citenamefont {{Mathews}}, \citenamefont {{Howard}}, \citenamefont
  {{Woosley}},\ and\ \citenamefont {{Hoffman}}}]{1992ApJ...399..656M}%
  \BibitemOpen
  \bibfield  {author} {\bibinfo {author} {\bibfnamefont {B.~S.}\ \bibnamefont
  {{Meyer}}}, \bibinfo {author} {\bibfnamefont {G.~J.}\ \bibnamefont
  {{Mathews}}}, \bibinfo {author} {\bibfnamefont {W.~M.}\ \bibnamefont
  {{Howard}}}, \bibinfo {author} {\bibfnamefont {S.~E.}\ \bibnamefont
  {{Woosley}}}, \ and\ \bibinfo {author} {\bibfnamefont {R.~D.}\ \bibnamefont
  {{Hoffman}}},\ }\href {\doibase 10.1086/171957} {\bibfield  {journal}
  {\bibinfo  {journal} {\apj}\ }\textbf {\bibinfo {volume} {399}},\ \bibinfo
  {pages} {656} (\bibinfo {year} {1992})}\BibitemShut {NoStop}%
\bibitem [{\citenamefont {Bliss}\ \emph {et~al.}(2017)\citenamefont {Bliss},
  \citenamefont {Arcones}, \citenamefont {Montes},\ and\ \citenamefont
  {Pereira}}]{Bliss}%
  \BibitemOpen
  \bibfield  {author} {\bibinfo {author} {\bibfnamefont {J.}~\bibnamefont
  {Bliss}}, \bibinfo {author} {\bibfnamefont {A.}~\bibnamefont {Arcones}},
  \bibinfo {author} {\bibfnamefont {F.}~\bibnamefont {Montes}}, \ and\ \bibinfo
  {author} {\bibfnamefont {J.}~\bibnamefont {Pereira}},\ }\href {\doibase
  https://doi.org/10.1088/1361-6471/aa63bd} {\bibfield  {journal} {\bibinfo
  {journal} {J. Phys. G: Nucl. Part. Phys.}\ }\textbf {\bibinfo {volume}
  {44}},\ \bibinfo {pages} {054003} (\bibinfo {year} {2017})}\BibitemShut
  {NoStop}%
\bibitem [{\citenamefont {Bliss}\ \emph {et~al.}(2018)\citenamefont {Bliss},
  \citenamefont {Witt}, \citenamefont {Arcones}, \citenamefont {Montes},\ and\
  \citenamefont {Pereira}}]{Bliss2018}%
  \BibitemOpen
  \bibfield  {author} {\bibinfo {author} {\bibfnamefont {J.}~\bibnamefont
  {Bliss}}, \bibinfo {author} {\bibfnamefont {M.}~\bibnamefont {Witt}},
  \bibinfo {author} {\bibfnamefont {A.}~\bibnamefont {Arcones}}, \bibinfo
  {author} {\bibfnamefont {F.}~\bibnamefont {Montes}}, \ and\ \bibinfo {author}
  {\bibfnamefont {J.}~\bibnamefont {Pereira}},\ }\href {\doibase
  https://doi.org/10.3847/1538-4357/aaadbe} {\bibfield  {journal} {\bibinfo
  {journal} {\apj}\ }\textbf {\bibinfo {volume} {855}},\ \bibinfo {pages} {135}
  (\bibinfo {year} {2018})}\BibitemShut {NoStop}%
\bibitem [{\citenamefont {Bliss}\ \emph {et~al.}(2020)\citenamefont {Bliss},
  \citenamefont {Arcones}, \citenamefont {Montes},\ and\ \citenamefont
  {Pereira}}]{bliss2020nuclear}%
  \BibitemOpen
  \bibfield  {author} {\bibinfo {author} {\bibfnamefont {J.}~\bibnamefont
  {Bliss}}, \bibinfo {author} {\bibfnamefont {A.}~\bibnamefont {Arcones}},
  \bibinfo {author} {\bibfnamefont {F.}~\bibnamefont {Montes}}, \ and\ \bibinfo
  {author} {\bibfnamefont {J.}~\bibnamefont {Pereira}},\ }\href {\doibase
  https://doi.org/10.1103/PhysRevC.101.055807} {\bibfield  {journal} {\bibinfo
  {journal} {\prc}\ }\textbf {\bibinfo {volume} {101}},\ \bibinfo {pages}
  {055807} (\bibinfo {year} {2020})}\BibitemShut {NoStop}%
\bibitem [{\citenamefont {Rauscher}\ \emph {et~al.}(1997)\citenamefont
  {Rauscher}, \citenamefont {Thielemann},\ and\ \citenamefont
  {Kratz}}]{PhysRevC.56.1613}%
  \BibitemOpen
  \bibfield  {author} {\bibinfo {author} {\bibfnamefont {T.}~\bibnamefont
  {Rauscher}}, \bibinfo {author} {\bibfnamefont {F.-K.}\ \bibnamefont
  {Thielemann}}, \ and\ \bibinfo {author} {\bibfnamefont {K.-L.}\ \bibnamefont
  {Kratz}},\ }\href {\doibase https://doi.org/10.1103/PhysRevC.56.1613}
  {\bibfield  {journal} {\bibinfo  {journal} {Phys. Rev. C}\ }\textbf {\bibinfo
  {volume} {56}},\ \bibinfo {pages} {1613} (\bibinfo {year}
  {1997})}\BibitemShut {NoStop}%
\bibitem [{\citenamefont {Mohr}\ \emph {et~al.}(2021)\citenamefont {Mohr},
  \citenamefont {Fülöp}, \citenamefont {Gyürky}, \citenamefont {Kiss},
  \citenamefont {Szücs}, \citenamefont {Arcones}, \citenamefont {Jacobi},\
  and\ \citenamefont {Psaltis}}]{atomkiv2}%
  \BibitemOpen
  \bibfield  {author} {\bibinfo {author} {\bibfnamefont {P.}~\bibnamefont
  {Mohr}}, \bibinfo {author} {\bibfnamefont {Z.}~\bibnamefont {Fülöp}},
  \bibinfo {author} {\bibfnamefont {G.}~\bibnamefont {Gyürky}}, \bibinfo
  {author} {\bibfnamefont {G.}~\bibnamefont {Kiss}}, \bibinfo {author}
  {\bibfnamefont {T.}~\bibnamefont {Szücs}}, \bibinfo {author} {\bibfnamefont
  {A.}~\bibnamefont {Arcones}}, \bibinfo {author} {\bibfnamefont
  {M.}~\bibnamefont {Jacobi}}, \ and\ \bibinfo {author} {\bibfnamefont
  {A.}~\bibnamefont {Psaltis}},\ }\href {\doibase
  https://doi.org/10.1016/j.adt.2021.101453} {\bibfield  {journal} {\bibinfo
  {journal} {Atomic Data and Nuclear Data Tables}\ }\textbf {\bibinfo {volume}
  {142}},\ \bibinfo {pages} {101453} (\bibinfo {year} {2021})}\BibitemShut
  {NoStop}%
\bibitem [{\citenamefont {Ong}\ \emph {et~al.}(2022)\citenamefont {Ong},
  \citenamefont {Avila}, \citenamefont {Mohr}, \citenamefont {Rehm},
  \citenamefont {Santiago-Gonzalez}, \citenamefont {Chen}, \citenamefont
  {Hoffman}, \citenamefont {Meisel}, \citenamefont {Montes},\ and\
  \citenamefont {Pereira}}]{Ong22}%
  \BibitemOpen
  \bibfield  {author} {\bibinfo {author} {\bibfnamefont {W.-J.}\ \bibnamefont
  {Ong}}, \bibinfo {author} {\bibfnamefont {M.~L.}\ \bibnamefont {Avila}},
  \bibinfo {author} {\bibfnamefont {P.}~\bibnamefont {Mohr}}, \bibinfo {author}
  {\bibfnamefont {K.~E.}\ \bibnamefont {Rehm}}, \bibinfo {author}
  {\bibfnamefont {D.}~\bibnamefont {Santiago-Gonzalez}}, \bibinfo {author}
  {\bibfnamefont {J.}~\bibnamefont {Chen}}, \bibinfo {author} {\bibfnamefont
  {C.~R.}\ \bibnamefont {Hoffman}}, \bibinfo {author} {\bibfnamefont
  {Z.}~\bibnamefont {Meisel}}, \bibinfo {author} {\bibfnamefont
  {F.}~\bibnamefont {Montes}}, \ and\ \bibinfo {author} {\bibfnamefont
  {J.}~\bibnamefont {Pereira}},\ }\href {\doibase 10.1103/PhysRevC.105.055803}
  {\bibfield  {journal} {\bibinfo  {journal} {Phys. Rev. C}\ }\textbf {\bibinfo
  {volume} {105}},\ \bibinfo {pages} {055803} (\bibinfo {year}
  {2022})}\BibitemShut {NoStop}%
\bibitem [{\citenamefont {Szegedi}\ \emph {et~al.}(2021)\citenamefont
  {Szegedi}, \citenamefont {Kiss}, \citenamefont {Mohr}, \citenamefont
  {Psaltis}, \citenamefont {Jacobi}, \citenamefont {Barnaf\"oldi},
  \citenamefont {Sz\"ucs}, \citenamefont {Gy\"urky},\ and\ \citenamefont
  {Arcones}}]{PhysRevC.104.035804}%
  \BibitemOpen
  \bibfield  {author} {\bibinfo {author} {\bibfnamefont {T.~N.}\ \bibnamefont
  {Szegedi}}, \bibinfo {author} {\bibfnamefont {G.~G.}\ \bibnamefont {Kiss}},
  \bibinfo {author} {\bibfnamefont {P.}~\bibnamefont {Mohr}}, \bibinfo {author}
  {\bibfnamefont {A.}~\bibnamefont {Psaltis}}, \bibinfo {author} {\bibfnamefont
  {M.}~\bibnamefont {Jacobi}}, \bibinfo {author} {\bibfnamefont {G.~G.}\
  \bibnamefont {Barnaf\"oldi}}, \bibinfo {author} {\bibfnamefont
  {T.}~\bibnamefont {Sz\"ucs}}, \bibinfo {author} {\bibfnamefont
  {G.}~\bibnamefont {Gy\"urky}}, \ and\ \bibinfo {author} {\bibfnamefont
  {A.}~\bibnamefont {Arcones}},\ }\href {\doibase 10.1103/PhysRevC.104.035804}
  {\bibfield  {journal} {\bibinfo  {journal} {Phys. Rev. C}\ }\textbf {\bibinfo
  {volume} {104}},\ \bibinfo {pages} {035804} (\bibinfo {year}
  {2021})}\BibitemShut {NoStop}%
\bibitem [{\citenamefont {Kiss}\ \emph {et~al.}(2021)\citenamefont {Kiss},
  \citenamefont {Szegedi}, \citenamefont {Mohr}, \citenamefont {Jacobi},
  \citenamefont {Gyürky}, \citenamefont {Huszánk},\ and\ \citenamefont
  {Arcones}}]{Kiss_2021}%
  \BibitemOpen
  \bibfield  {author} {\bibinfo {author} {\bibfnamefont {G.~G.}\ \bibnamefont
  {Kiss}}, \bibinfo {author} {\bibfnamefont {T.~N.}\ \bibnamefont {Szegedi}},
  \bibinfo {author} {\bibfnamefont {P.}~\bibnamefont {Mohr}}, \bibinfo {author}
  {\bibfnamefont {M.}~\bibnamefont {Jacobi}}, \bibinfo {author} {\bibfnamefont
  {G.}~\bibnamefont {Gyürky}}, \bibinfo {author} {\bibfnamefont
  {R.}~\bibnamefont {Huszánk}}, \ and\ \bibinfo {author} {\bibfnamefont
  {A.}~\bibnamefont {Arcones}},\ }\href {\doibase 10.3847/1538-4357/abd2bc}
  {\bibfield  {journal} {\bibinfo  {journal} {\apj}\ }\textbf {\bibinfo
  {volume} {908}},\ \bibinfo {pages} {202} (\bibinfo {year}
  {2021})}\BibitemShut {NoStop}%
\bibitem [{\citenamefont {Carnelli}\ \emph {et~al.}(2015)\citenamefont
  {Carnelli}, \citenamefont {Almaraz-Calderon}, \citenamefont {Rehm},
  \citenamefont {Albers}, \citenamefont {Alcorta}, \citenamefont {Bertone},
  \citenamefont {Digiovine}, \citenamefont {Esbensen}, \citenamefont
  {{Fernández Niello}}, \citenamefont {Henderson} \emph
  {et~al.}}]{MUSICsetup}%
  \BibitemOpen
  \bibfield  {author} {\bibinfo {author} {\bibfnamefont {P.}~\bibnamefont
  {Carnelli}}, \bibinfo {author} {\bibfnamefont {S.}~\bibnamefont
  {Almaraz-Calderon}}, \bibinfo {author} {\bibfnamefont {K.}~\bibnamefont
  {Rehm}}, \bibinfo {author} {\bibfnamefont {M.}~\bibnamefont {Albers}},
  \bibinfo {author} {\bibfnamefont {M.}~\bibnamefont {Alcorta}}, \bibinfo
  {author} {\bibfnamefont {P.}~\bibnamefont {Bertone}}, \bibinfo {author}
  {\bibfnamefont {B.}~\bibnamefont {Digiovine}}, \bibinfo {author}
  {\bibfnamefont {H.}~\bibnamefont {Esbensen}}, \bibinfo {author}
  {\bibfnamefont {J.}~\bibnamefont {{Fernández Niello}}}, \bibinfo {author}
  {\bibfnamefont {D.}~\bibnamefont {Henderson}},  \emph {et~al.},\ }\href
  {\doibase https://doi.org/10.1016/j.nima.2015.07.030} {\bibfield  {journal}
  {\bibinfo  {journal} {Nucl. Inst. and Meth. in Phys. Res. Section A}\
  }\textbf {\bibinfo {volume} {799}},\ \bibinfo {pages} {197} (\bibinfo {year}
  {2015})}\BibitemShut {NoStop}%
\bibitem [{\citenamefont {Avila}\ \emph {et~al.}(2017)\citenamefont {Avila},
  \citenamefont {Rehm}, \citenamefont {Almaraz-Calderon}, \citenamefont
  {Ayangeakaa}, \citenamefont {Dickerson}, \citenamefont {Hoffman},
  \citenamefont {Jiang}, \citenamefont {Kay}, \citenamefont {Lai},
  \citenamefont {Nusair} \emph {et~al.}}]{AvilaNIM16}%
  \BibitemOpen
  \bibfield  {author} {\bibinfo {author} {\bibfnamefont {M.~L.}\ \bibnamefont
  {Avila}}, \bibinfo {author} {\bibfnamefont {K.~E.}\ \bibnamefont {Rehm}},
  \bibinfo {author} {\bibfnamefont {S.}~\bibnamefont {Almaraz-Calderon}},
  \bibinfo {author} {\bibfnamefont {A.~D.}\ \bibnamefont {Ayangeakaa}},
  \bibinfo {author} {\bibfnamefont {C.}~\bibnamefont {Dickerson}}, \bibinfo
  {author} {\bibfnamefont {C.~R.}\ \bibnamefont {Hoffman}}, \bibinfo {author}
  {\bibfnamefont {C.~L.}\ \bibnamefont {Jiang}}, \bibinfo {author}
  {\bibfnamefont {B.~P.}\ \bibnamefont {Kay}}, \bibinfo {author} {\bibfnamefont
  {J.}~\bibnamefont {Lai}}, \bibinfo {author} {\bibfnamefont {O.}~\bibnamefont
  {Nusair}},  \emph {et~al.},\ }\href {\doibase 10.1016/j.nima.2017.03.060}
  {\bibfield  {journal} {\bibinfo  {journal} {Nucl. Inst. and Meth. in Phys.
  Res. Section A}\ }\textbf {\bibinfo {volume} {859}},\ \bibinfo {pages} {63}
  (\bibinfo {year} {2017})}\BibitemShut {NoStop}%
\bibitem [{\citenamefont {Avila}\ \emph {et~al.}(2016)\citenamefont {Avila},
  \citenamefont {Rehm}, \citenamefont {Almaraz-Calderon}, \citenamefont
  {Ayangeakaa}, \citenamefont {Dickerson}, \citenamefont {Hoffman},
  \citenamefont {Jiang}, \citenamefont {Kay}, \citenamefont {Lai},
  \citenamefont {Nusair} \emph {et~al.}}]{Avila16}%
  \BibitemOpen
  \bibfield  {author} {\bibinfo {author} {\bibfnamefont {M.~L.}\ \bibnamefont
  {Avila}}, \bibinfo {author} {\bibfnamefont {K.~E.}\ \bibnamefont {Rehm}},
  \bibinfo {author} {\bibfnamefont {S.}~\bibnamefont {Almaraz-Calderon}},
  \bibinfo {author} {\bibfnamefont {A.~D.}\ \bibnamefont {Ayangeakaa}},
  \bibinfo {author} {\bibfnamefont {C.}~\bibnamefont {Dickerson}}, \bibinfo
  {author} {\bibfnamefont {C.~R.}\ \bibnamefont {Hoffman}}, \bibinfo {author}
  {\bibfnamefont {C.~L.}\ \bibnamefont {Jiang}}, \bibinfo {author}
  {\bibfnamefont {B.~P.}\ \bibnamefont {Kay}}, \bibinfo {author} {\bibfnamefont
  {J.}~\bibnamefont {Lai}}, \bibinfo {author} {\bibfnamefont {O.}~\bibnamefont
  {Nusair}},  \emph {et~al.},\ }\href {\doibase 10.1103/PhysRevC.94.065804}
  {\bibfield  {journal} {\bibinfo  {journal} {Phys. Rev. C}\ }\textbf {\bibinfo
  {volume} {94}},\ \bibinfo {pages} {065804} (\bibinfo {year}
  {2016})}\BibitemShut {NoStop}%
\bibitem [{\citenamefont {Weick}\ \emph {et~al.}(2018)\citenamefont {Weick},
  \citenamefont {Geissel}, \citenamefont {Iwasa}, \citenamefont
  {Scheidenberger}, \citenamefont {Sanchez}, \citenamefont {Prochazka},\ and\
  \citenamefont {Purushotaman}}]{ATIMA}%
  \BibitemOpen
  \bibfield  {author} {\bibinfo {author} {\bibfnamefont {H.}~\bibnamefont
  {Weick}}, \bibinfo {author} {\bibfnamefont {H.}~\bibnamefont {Geissel}},
  \bibinfo {author} {\bibfnamefont {N.}~\bibnamefont {Iwasa}}, \bibinfo
  {author} {\bibfnamefont {C.}~\bibnamefont {Scheidenberger}}, \bibinfo
  {author} {\bibfnamefont {J.~R.}\ \bibnamefont {Sanchez}}, \bibinfo {author}
  {\bibfnamefont {A.}~\bibnamefont {Prochazka}}, \ and\ \bibinfo {author}
  {\bibfnamefont {S.}~\bibnamefont {Purushotaman}},\ }\href@noop {} {\bibfield
  {journal} {\bibinfo  {journal} {GSI Sci. Rep.}\ }\textbf {\bibinfo {volume}
  {2018--1}},\ \bibinfo {pages} {130} (\bibinfo {year} {2018})}\BibitemShut
  {NoStop}%
\bibitem [{\citenamefont {{ Ziegler}}\ \emph {et~al.}(2013)\citenamefont {{
  Ziegler}}, \citenamefont {Ziegler},\ and\ \citenamefont {Biersack}}]{srim}%
  \BibitemOpen
  \bibfield  {author} {\bibinfo {author} {\bibfnamefont {J.~F.}\ \bibnamefont
  {{ Ziegler}}}, \bibinfo {author} {\bibfnamefont {M.~D.}\ \bibnamefont
  {Ziegler}}, \ and\ \bibinfo {author} {\bibfnamefont {J.~P.}\ \bibnamefont
  {Biersack}},\ }\href@noop {} {\bibfield  {journal} {\bibinfo  {journal} {SRIM
  Co., United States of America}\ }\textbf {\bibinfo {volume} {6$^{th}$}}
  (\bibinfo {year} {2013})}\BibitemShut {NoStop}%
\bibitem [{\citenamefont {Wilkins}\ \emph {et~al.}(1971)\citenamefont
  {Wilkins}, \citenamefont {Fluss}, \citenamefont {Kaufman}, \citenamefont
  {Gross},\ and\ \citenamefont {Steinberg}}]{WILKINS1971381}%
  \BibitemOpen
  \bibfield  {author} {\bibinfo {author} {\bibfnamefont {B.}~\bibnamefont
  {Wilkins}}, \bibinfo {author} {\bibfnamefont {M.}~\bibnamefont {Fluss}},
  \bibinfo {author} {\bibfnamefont {S.}~\bibnamefont {Kaufman}}, \bibinfo
  {author} {\bibfnamefont {C.}~\bibnamefont {Gross}}, \ and\ \bibinfo {author}
  {\bibfnamefont {E.}~\bibnamefont {Steinberg}},\ }\href {\doibase
  https://doi.org/10.1016/0029-554X(71)90414-9} {\bibfield  {journal} {\bibinfo
   {journal} {Nucl. Inst. and Meth.}\ }\textbf {\bibinfo {volume} {92}},\
  \bibinfo {pages} {381} (\bibinfo {year} {1971})}\BibitemShut {NoStop}%
\bibitem [{\citenamefont {Tarasov}\ and\ \citenamefont {Bazin}(2008)}]{lise}%
  \BibitemOpen
  \bibfield  {author} {\bibinfo {author} {\bibfnamefont {O.}~\bibnamefont
  {Tarasov}}\ and\ \bibinfo {author} {\bibfnamefont {D.}~\bibnamefont
  {Bazin}},\ }\href {\doibase https://doi.org/10.1016/j.nimb.2008.05.110}
  {\bibfield  {journal} {\bibinfo  {journal} {Nucl. Instr. and Meth. in Phys.
  Res. B}\ }\textbf {\bibinfo {volume} {266}},\ \bibinfo {pages} {4657}
  (\bibinfo {year} {2008})}\BibitemShut {NoStop}%
\bibitem [{\citenamefont {Koning}\ \emph {et~al.}(2019)\citenamefont {Koning},
  \citenamefont {Rochman}, \citenamefont {Sublet}, \citenamefont {Dzysiuk},
  \citenamefont {Fleming},\ and\ \citenamefont {van~der Marck}}]{talys1}%
  \BibitemOpen
  \bibfield  {author} {\bibinfo {author} {\bibfnamefont {A.}~\bibnamefont
  {Koning}}, \bibinfo {author} {\bibfnamefont {D.}~\bibnamefont {Rochman}},
  \bibinfo {author} {\bibfnamefont {J.-C.}\ \bibnamefont {Sublet}}, \bibinfo
  {author} {\bibfnamefont {N.}~\bibnamefont {Dzysiuk}}, \bibinfo {author}
  {\bibfnamefont {M.}~\bibnamefont {Fleming}}, \ and\ \bibinfo {author}
  {\bibfnamefont {S.}~\bibnamefont {van~der Marck}},\ }\href {\doibase
  https://doi.org/10.1016/j.nds.2019.01.002} {\bibfield  {journal} {\bibinfo
  {journal} {Nuclear Data Sheets}\ }\textbf {\bibinfo {volume} {155}},\
  \bibinfo {pages} {1} (\bibinfo {year} {2019})}\BibitemShut {NoStop}%
\bibitem [{\citenamefont {Koning}\ and\ \citenamefont
  {Rochman}(2012)}]{talys2}%
  \BibitemOpen
  \bibfield  {author} {\bibinfo {author} {\bibfnamefont {A.}~\bibnamefont
  {Koning}}\ and\ \bibinfo {author} {\bibfnamefont {D.}~\bibnamefont
  {Rochman}},\ }\href {\doibase https://doi.org/10.1016/j.nds.2012.11.002}
  {\bibfield  {journal} {\bibinfo  {journal} {Nuclear Data Sheets}\ }\textbf
  {\bibinfo {volume} {113}},\ \bibinfo {pages} {2841} (\bibinfo {year}
  {2012})}\BibitemShut {NoStop}%
\bibitem [{\citenamefont {Workman}\ \emph {et~al.}(2022)\citenamefont {Workman}
  \emph {et~al.}}]{bookPDG}%
  \BibitemOpen
  \bibfield  {author} {\bibinfo {author} {\bibfnamefont {R.~L.}\ \bibnamefont
  {Workman}} \emph {et~al.},\ }\href@noop {} {\emph {\bibinfo {title} {Particle
  Data Group. Review of {P}article {P}hysics.}}}\ (\bibinfo  {publisher}
  {PTEPr},\ \bibinfo {year} {2022})\BibitemShut {NoStop}%
\bibitem [{\citenamefont {Avrigeanu}\ \emph {et~al.}(2014)\citenamefont
  {Avrigeanu}, \citenamefont {Avrigeanu},\ and\ \citenamefont {M\ifmmode
  \u{a}\else \u{a}\fi{}n\ifmmode~\u{a}\else \u{a}\fi{}ilescu}}]{Avrigeanu2014}%
  \BibitemOpen
  \bibfield  {author} {\bibinfo {author} {\bibfnamefont {V.}~\bibnamefont
  {Avrigeanu}}, \bibinfo {author} {\bibfnamefont {M.}~\bibnamefont
  {Avrigeanu}}, \ and\ \bibinfo {author} {\bibfnamefont {C.}~\bibnamefont
  {M\ifmmode \u{a}\else \u{a}\fi{}n\ifmmode~\u{a}\else \u{a}\fi{}ilescu}},\
  }\href {\doibase https://doi.org/10.1103/PhysRevC.90.044612} {\bibfield
  {journal} {\bibinfo  {journal} {Phys. Rev. C}\ }\textbf {\bibinfo {volume}
  {90}},\ \bibinfo {pages} {044612} (\bibinfo {year} {2014})}\BibitemShut
  {NoStop}%
\bibitem [{\citenamefont {Avrigeanu}\ and\ \citenamefont
  {Avrigeanu}(2023)}]{Avrigeanu2023}%
  \BibitemOpen
  \bibfield  {author} {\bibinfo {author} {\bibfnamefont {M.}~\bibnamefont
  {Avrigeanu}}\ and\ \bibinfo {author} {\bibfnamefont {V.}~\bibnamefont
  {Avrigeanu}},\ }\href {\doibase https://doi.org/10.1103/PhysRevC.107.034613}
  {\bibfield  {journal} {\bibinfo  {journal} {Phys. Rev. C}\ }\textbf {\bibinfo
  {volume} {107}},\ \bibinfo {pages} {034613} (\bibinfo {year}
  {2023})}\BibitemShut {NoStop}%
\bibitem [{\citenamefont {Oprea}\ \emph {et~al.}(2017)\citenamefont {Oprea},
  \citenamefont {Glodariu}, \citenamefont {Filipescu}, \citenamefont
  {Gheorghe}, \citenamefont {Mitu}, \citenamefont {Boromiza}, \citenamefont
  {Bucurescu}, \citenamefont {Costache}, \citenamefont {Cata-Danil},
  \citenamefont {Florea} \emph {et~al.}}]{ref86Sr}%
  \BibitemOpen
  \bibfield  {author} {\bibinfo {author} {\bibfnamefont {A.}~\bibnamefont
  {Oprea}}, \bibinfo {author} {\bibfnamefont {T.}~\bibnamefont {Glodariu}},
  \bibinfo {author} {\bibfnamefont {D.}~\bibnamefont {Filipescu}}, \bibinfo
  {author} {\bibfnamefont {I.}~\bibnamefont {Gheorghe}}, \bibinfo {author}
  {\bibfnamefont {A.}~\bibnamefont {Mitu}}, \bibinfo {author} {\bibfnamefont
  {M.}~\bibnamefont {Boromiza}}, \bibinfo {author} {\bibfnamefont
  {D.}~\bibnamefont {Bucurescu}}, \bibinfo {author} {\bibfnamefont
  {C.}~\bibnamefont {Costache}}, \bibinfo {author} {\bibfnamefont
  {I.}~\bibnamefont {Cata-Danil}}, \bibinfo {author} {\bibfnamefont
  {N.}~\bibnamefont {Florea}},  \emph {et~al.},\ }\href {\doibase
  https://doi.org/10.1051/epjconf/201714601016} {\bibfield  {journal} {\bibinfo
   {journal} {EPJ Web Conf.}\ }\textbf {\bibinfo {volume} {146}},\ \bibinfo
  {pages} {01016} (\bibinfo {year} {2017})}\BibitemShut {NoStop}%
\bibitem [{\citenamefont {Rapp}\ \emph {et~al.}(2008)\citenamefont {Rapp},
  \citenamefont {Dillmann}, \citenamefont {K\"appeler}, \citenamefont {Giesen},
  \citenamefont {Klein}, \citenamefont {Rauscher}, \citenamefont {Hentschel},\
  and\ \citenamefont {Hilpp}}]{PhysRevC.78.025804}%
  \BibitemOpen
  \bibfield  {author} {\bibinfo {author} {\bibfnamefont {W.}~\bibnamefont
  {Rapp}}, \bibinfo {author} {\bibfnamefont {I.}~\bibnamefont {Dillmann}},
  \bibinfo {author} {\bibfnamefont {F.}~\bibnamefont {K\"appeler}}, \bibinfo
  {author} {\bibfnamefont {U.}~\bibnamefont {Giesen}}, \bibinfo {author}
  {\bibfnamefont {H.}~\bibnamefont {Klein}}, \bibinfo {author} {\bibfnamefont
  {T.}~\bibnamefont {Rauscher}}, \bibinfo {author} {\bibfnamefont
  {D.}~\bibnamefont {Hentschel}}, \ and\ \bibinfo {author} {\bibfnamefont
  {S.}~\bibnamefont {Hilpp}},\ }\href {\doibase 10.1103/PhysRevC.78.025804}
  {\bibfield  {journal} {\bibinfo  {journal} {Phys. Rev. C}\ }\textbf {\bibinfo
  {volume} {78}},\ \bibinfo {pages} {025804} (\bibinfo {year}
  {2008})}\BibitemShut {NoStop}%
\bibitem [{\citenamefont {Jayatissa}\ \emph {et~al.}(2022)\citenamefont
  {Jayatissa}, \citenamefont {Avila}, \citenamefont {Rehm}, \citenamefont
  {Talwar}, \citenamefont {Mohr}, \citenamefont {Auranen}, \citenamefont
  {Chen}, \citenamefont {Gorelov}, \citenamefont {Hoffman}, \citenamefont
  {Jiang} \emph {et~al.}}]{PhysRevC.105.L042802}%
  \BibitemOpen
  \bibfield  {author} {\bibinfo {author} {\bibfnamefont {H.}~\bibnamefont
  {Jayatissa}}, \bibinfo {author} {\bibfnamefont {M.~L.}\ \bibnamefont
  {Avila}}, \bibinfo {author} {\bibfnamefont {K.~E.}\ \bibnamefont {Rehm}},
  \bibinfo {author} {\bibfnamefont {R.}~\bibnamefont {Talwar}}, \bibinfo
  {author} {\bibfnamefont {P.}~\bibnamefont {Mohr}}, \bibinfo {author}
  {\bibfnamefont {K.}~\bibnamefont {Auranen}}, \bibinfo {author} {\bibfnamefont
  {J.}~\bibnamefont {Chen}}, \bibinfo {author} {\bibfnamefont {D.~A.}\
  \bibnamefont {Gorelov}}, \bibinfo {author} {\bibfnamefont {C.~R.}\
  \bibnamefont {Hoffman}}, \bibinfo {author} {\bibfnamefont {C.~L.}\
  \bibnamefont {Jiang}},  \emph {et~al.},\ }\href {\doibase
  https://doi.org/10.1103/PhysRevC.105.L042802} {\bibfield  {journal} {\bibinfo
   {journal} {Phys. Rev. C}\ }\textbf {\bibinfo {volume} {105}},\ \bibinfo
  {pages} {L042802} (\bibinfo {year} {2022})}\BibitemShut {NoStop}%
\bibitem [{\citenamefont {Koning}\ \emph {et~al.}(2008)\citenamefont {Koning},
  \citenamefont {Hilaire},\ and\ \citenamefont {Duijvestijn}}]{talys_general}%
  \BibitemOpen
  \bibfield  {author} {\bibinfo {author} {\bibfnamefont {A.}~\bibnamefont
  {Koning}}, \bibinfo {author} {\bibfnamefont {S.}~\bibnamefont {Hilaire}}, \
  and\ \bibinfo {author} {\bibfnamefont {M.}~\bibnamefont {Duijvestijn}},\ }in\
  \href@noop {} {\emph {\bibinfo {booktitle} {Proceedings of the International
  Conference on Nuclear Data for Science and Technology}}},\ \bibinfo {editor}
  {edited by\ \bibinfo {editor} {\bibnamefont {O.Bersillon}}, \bibinfo {editor}
  {\bibnamefont {F.Gunsing}}, \bibinfo {editor} {\bibnamefont {E.Bauge}}, \
  and\ \bibinfo {editor} {\bibnamefont {R.Jacqmin}}}\ (\bibinfo  {publisher}
  {EDP Sciences},\ \bibinfo {address} {April 22-27, 2007, Nice, France},\
  \bibinfo {year} {2008})\ pp.\ \bibinfo {pages} {211--214}\BibitemShut
  {NoStop}%
\bibitem [{\citenamefont {{De Vries}}\ \emph {et~al.}(1987)\citenamefont {{De
  Vries}}, \citenamefont {{De Jager}},\ and\ \citenamefont {{De
  Vries}}}]{DEVRIES1987495}%
  \BibitemOpen
  \bibfield  {author} {\bibinfo {author} {\bibfnamefont {H.}~\bibnamefont {{De
  Vries}}}, \bibinfo {author} {\bibfnamefont {C.}~\bibnamefont {{De Jager}}}, \
  and\ \bibinfo {author} {\bibfnamefont {C.}~\bibnamefont {{De Vries}}},\
  }\href {\doibase 10.1016/0092-640X(87)90013-1} {\bibfield  {journal}
  {\bibinfo  {journal} {Atomic Data and Nuclear Data Tables}\ }\textbf
  {\bibinfo {volume} {36}},\ \bibinfo {pages} {495} (\bibinfo {year}
  {1987})}\BibitemShut {NoStop}%
\bibitem [{\citenamefont {Mohr}\ \emph {et~al.}(2013)\citenamefont {Mohr},
  \citenamefont {Kiss}, \citenamefont {Fülöp}, \citenamefont {Galaviz},
  \citenamefont {Gyürky},\ and\ \citenamefont {Somorjai}}]{redSigma}%
  \BibitemOpen
  \bibfield  {author} {\bibinfo {author} {\bibfnamefont {P.}~\bibnamefont
  {Mohr}}, \bibinfo {author} {\bibfnamefont {G.}~\bibnamefont {Kiss}}, \bibinfo
  {author} {\bibfnamefont {Z.}~\bibnamefont {Fülöp}}, \bibinfo {author}
  {\bibfnamefont {D.}~\bibnamefont {Galaviz}}, \bibinfo {author} {\bibfnamefont
  {G.}~\bibnamefont {Gyürky}}, \ and\ \bibinfo {author} {\bibfnamefont
  {E.}~\bibnamefont {Somorjai}},\ }\href {\doibase
  https://doi.org/10.1016/j.adt.2012.10.003} {\bibfield  {journal} {\bibinfo
  {journal} {Atomic Data and Nuclear Data Tables}\ }\textbf {\bibinfo {volume}
  {99}},\ \bibinfo {pages} {651} (\bibinfo {year} {2013})}\BibitemShut
  {NoStop}%
\bibitem [{\citenamefont {Gomes}\ \emph {et~al.}(2005)\citenamefont {Gomes},
  \citenamefont {Lubian}, \citenamefont {Padron},\ and\ \citenamefont
  {Anjos}}]{Gomes_PRC2005_reduced}%
  \BibitemOpen
  \bibfield  {author} {\bibinfo {author} {\bibfnamefont {P.~R.~S.}\
  \bibnamefont {Gomes}}, \bibinfo {author} {\bibfnamefont {J.}~\bibnamefont
  {Lubian}}, \bibinfo {author} {\bibfnamefont {I.}~\bibnamefont {Padron}}, \
  and\ \bibinfo {author} {\bibfnamefont {R.~M.}\ \bibnamefont {Anjos}},\ }\href
  {\doibase https://doi.org/10.1103/PhysRevC.71.017601} {\bibfield  {journal}
  {\bibinfo  {journal} {Phys. Rev. C}\ }\textbf {\bibinfo {volume} {71}},\
  \bibinfo {pages} {017601} (\bibinfo {year} {2005})}\BibitemShut {NoStop}%
\bibitem [{\citenamefont {Datta}\ \emph {et~al.}(1989)\citenamefont {Datta},
  \citenamefont {Ray}, \citenamefont {Majumdar}, \citenamefont {Ghosh},
  \citenamefont {Samanta}, \citenamefont {Ray}, \citenamefont {Dasgupta},
  \citenamefont {Chintalapudi},\ and\ \citenamefont
  {Banerjee}}]{PhysRevC.39.1281}%
  \BibitemOpen
  \bibfield  {author} {\bibinfo {author} {\bibfnamefont {S.~K.}\ \bibnamefont
  {Datta}}, \bibinfo {author} {\bibfnamefont {S.}~\bibnamefont {Ray}}, \bibinfo
  {author} {\bibfnamefont {H.}~\bibnamefont {Majumdar}}, \bibinfo {author}
  {\bibfnamefont {S.~K.}\ \bibnamefont {Ghosh}}, \bibinfo {author}
  {\bibfnamefont {C.}~\bibnamefont {Samanta}}, \bibinfo {author} {\bibfnamefont
  {S.}~\bibnamefont {Ray}}, \bibinfo {author} {\bibfnamefont {P.}~\bibnamefont
  {Dasgupta}}, \bibinfo {author} {\bibfnamefont {S.~N.}\ \bibnamefont
  {Chintalapudi}}, \ and\ \bibinfo {author} {\bibfnamefont {S.~R.}\
  \bibnamefont {Banerjee}},\ }\href {\doibase
  https://doi.org/10.1103/PhysRevC.39.1281} {\bibfield  {journal} {\bibinfo
  {journal} {Phys. Rev. C}\ }\textbf {\bibinfo {volume} {39}},\ \bibinfo
  {pages} {1281} (\bibinfo {year} {1989})}\BibitemShut {NoStop}%
\bibitem [{\citenamefont {Rauscher}(2023)}]{refExp2Rate}%
  \BibitemOpen
  \bibfield  {author} {\bibinfo {author} {\bibfnamefont {T.}~\bibnamefont
  {Rauscher}},\ }\href@noop {} {\enquote {\bibinfo {title} {{{\tt EXP2RATE}
  V2.1} \url{http://nucastro.org/codes.html}},}\ } (\bibinfo {year}
  {2023})\BibitemShut {NoStop}%
\bibitem [{\citenamefont {Cyburt}\ \emph {et~al.}(2010)\citenamefont {Cyburt},
  \citenamefont {Amthor}, \citenamefont {Ferguson}, \citenamefont {Meisel},
  \citenamefont {Smith}, \citenamefont {Warren}, \citenamefont {Heger},
  \citenamefont {Hoffman}, \citenamefont {Rauscher}, \citenamefont {Sakharuk}
  \emph {et~al.}}]{Cyburt2010}%
  \BibitemOpen
  \bibfield  {author} {\bibinfo {author} {\bibfnamefont {R.~H.}\ \bibnamefont
  {Cyburt}}, \bibinfo {author} {\bibfnamefont {A.~M.}\ \bibnamefont {Amthor}},
  \bibinfo {author} {\bibfnamefont {R.}~\bibnamefont {Ferguson}}, \bibinfo
  {author} {\bibfnamefont {Z.}~\bibnamefont {Meisel}}, \bibinfo {author}
  {\bibfnamefont {K.}~\bibnamefont {Smith}}, \bibinfo {author} {\bibfnamefont
  {S.}~\bibnamefont {Warren}}, \bibinfo {author} {\bibfnamefont
  {A.}~\bibnamefont {Heger}}, \bibinfo {author} {\bibfnamefont {R.~D.}\
  \bibnamefont {Hoffman}}, \bibinfo {author} {\bibfnamefont {T.}~\bibnamefont
  {Rauscher}}, \bibinfo {author} {\bibfnamefont {A.}~\bibnamefont {Sakharuk}},
  \emph {et~al.},\ }\href {\doibase 10.1088/0067-0049/189/1/240} {\bibfield
  {journal} {\bibinfo  {journal} {Astrophys. J. Suppl. Ser.}\ }\textbf
  {\bibinfo {volume} {189}},\ \bibinfo {pages} {240} (\bibinfo {year}
  {2010})}\BibitemShut {NoStop}%
\end{thebibliography}%

\end{document}